\documentclass[twocolumn]{emulateapj}

\lefthead{Law and Majewski 2010}
\righthead{Evolution of Sgr in a Triaxial Halo}

\slugcomment{DRAFT: \today}

\begin{document}

\newcommand{\msun}{\ensuremath{\rm M_\odot}}
\newcommand{\msunyr}{\ensuremath{\rm M_{\odot}\;{\rm yr}^{-1}}}
\newcommand{\Ha}{\ensuremath{\rm H\alpha}}
\newcommand{\Hb}{\ensuremath{\rm H\beta}}
\newcommand{\lya}{\ensuremath{\rm Ly\alpha}}
\newcommand{\Ntwo}{[\ion{N}{2}]}
\newcommand{\kms}{\textrm{km~s}\ensuremath{^{-1}\,}}
\newcommand{\ztwo}{\ensuremath{z\sim2}}
\newcommand{\zthree}{\ensuremath{z\sim3}}
\newcommand{\feh}{\textrm{[Fe/H]}}

\newcommand{\hst}{{\it HST}-ACS}

\title{The Sagittarius Dwarf Galaxy: a Model for Evolution in a Triaxial Milky Way Halo}
\author{David R.~Law\altaffilmark{1,2} and Steven R. Majewski\altaffilmark{3}}

\altaffiltext{1}{Hubble Fellow.}
\altaffiltext{2}{Department of Physics and Astronomy, University of California, Los Angeles, CA 90095;
drlaw@astro.ucla.edu}
\altaffiltext{3}{Dept. of Astronomy, University of Virginia,
Charlottesville, VA 22904-4325: srm4n@virginia.edu}

\begin{abstract}

We present a new $N$-body model for the tidal disruption of the Sagittarius (Sgr) dwarf that is capable of simultaneously satisfying the majority of
angular position, distance, and radial velocity constraints imposed by current wide-field surveys of its dynamically young
($\lesssim 3$ Gyr) tidal debris streams.
In particular, this model resolves the conflicting angular position and radial velocity constraints on the Sgr leading tidal stream
that have been highlighted in recent years.
While the model does not reproduce the apparent bifurcation observed in the leading debris stream, recent observational
data suggest that this bifurcation may represent a constraint on the internal properties of the Sgr dwarf rather
than the details of its orbit.
The key element in the success of this model is the introduction
of a non-axisymmetric component to the Galactic gravitational potential which can be described in terms of a
triaxial  dark matter halo 
whose minor/major axis ratio $(c/a)_{\Phi} = 0.72$ and intermediate/major axis ratio $(b/a)_{\Phi} = 0.99$
at radii $20 < r < 60$ kpc.
The minor/intermediate/major axes of this halo lie along the directions
$(l, b) = (7^{\circ}, 0^{\circ})$, $(0^{\circ}, 90^{\circ})$, and $(97^{\circ}, 0^{\circ})$ respectively, 
corresponding to
a nearly-oblate ellipsoid  whose minor axis is contained within the Galactic disk plane.
This particular disk/halo orientation is difficult to reconcile within the general context of galactic dynamics (and CDM models in particular), 
suggesting either that the orientation may have evolved significantly with time or that inclusion of other non-axisymmetric components (such as the 
gravitational influence of the Magellanic Clouds) in the model may obviate the need for triaxiality in the dark matter halo.
The apparent proper motion of Sgr in this model is estimated to be $(\mu_l \textrm{cos}\, b, \mu_b) = (-2.16, 1.73)$ mas yr$^{-1}$,
corresponding to a Galactocentric space velocity $(U,V,W$) = (230, -35, 195) \kms.
Based on the velocity dispersion in the stellar tidal streams, we estimate that Sgr has a current  bound mass $M_{\rm Sgr} = 2.5^{+1.3}_{-1.0} \times 10^{8} M_{\odot}$.
We demonstrate that with simple assumptions about the star formation history of Sgr, tidal stripping models naturally give rise to gradients
in the metallicity distribution function (MDF) along the stellar debris streams similar to those observed in recent studies.  These models predict
a strong evolution in the MDF of the model Sgr dwarf with time, indicating that the chemical abundances of stars in Sgr at the present day may
be significantly different than the abundances of those already contributed to the Galactic stellar halo. 
We conclude by using the new $N$-body model to reevaluate previous claims of the association of miscellaneous halo substructure with the Sgr dwarf.

\end{abstract}


\section{INTRODUCTION}

With the advent of progressively deeper photometric surveys, it has become apparent that galactic haloes are threaded 
with the phase-mixed detritus of multiple generations of
dwarf satellites that have been destroyed by the inexorable tides of their host's gravitational potential.
Surveys have indicated the presence of such streams in both the Milky Way (e.g., Majewski et al. 1996; Helmi et al. 1999; Newberg et al. 2002; 
Rocha-Pinto et al. 2004; Belokurov et al. 2006; Duffau et al. 2006; Grillmair \& Dionatos 2006; Starkenburg et al. 2009) and 
other nearby galaxies (e.g., Shang et al. 1998; Ibata et al. 2001a; McConnachie et al. 2003; Kalirai et al. 2006;
Mart{\'{\i}}nez-Delgado et al. 2008, 2009), possibly constituting the primary source
of the Galactic stellar halo (e.g., Searle \& Zinn 1978; Majewski 1993, 2004; Majewski et al. 1996; Bullock et al. 2001b; Bullock \& Johnston 2005; Font et al. 2006; Bell et al. 2008).

Perhaps the most prominent such stream is that from the Sagittarius dwarf spheroidal galaxy 
(hereafter Sgr dSph), whose lengthy tidal streams wrap entirely around the Milky Way.  The first observations
of Sgr were presented by Ibata et al. (1994), since which time significant observational effort has been invested in detecting and characterizing the stellar streams emanating from the dwarf.
Observations of high-latitude carbon stars (Totten \& Irwin 1998) and various small-field surveys (e.g., Dinescu et al. 2002, and references therein) fueled a host of early efforts to model the
Sgr --- Milky Way system by Johnston et al. (1995, 1999), Velazquez \& White (1995), Edelsohn \& Elmegreen (1997),
Ibata et al. (1997), G{\'o}mez-Flechoso et al. (1999), Helmi \& White (2001), Mart{\'{\i}}nez-Delgado et al. (2004).

In the past seven years however, our understanding of the scope and significance of the streams has been revolutionized by the deep, wide-field views provided by 
the Two Micron All-Sky Survey (2MASS) and Sloan Digital Sky Survey (SDSS).
In particular,
the 2MASS survey showed a large population of young, relatively metal-rich Sgr M-giants wrapping $360^{\circ}$ or more across the sky (Majewski et al. 2003).
SDSS observations have shown (Belokurov et al. 2006) that 
the debris streamer
leading Sgr along its orbit continues to be well-defined through the North Galactic Pole as it passes over the solar neighborhood towards the Galactic anticenter.
Follow-up spectroscopy (Majewski et al. 2004; Law et al. 2004; Yanny et al. 2009) has confirmed that the streams are similarly well-defined in radial velocity, and indicate a significant
evolution in the metallicity distribution function (MDF) of the component stellar populations with increasing separation from the Sgr dwarf core along the tidal streams
(e.g., Bellazini et al. 2006; Chou et al. 2007, 2009; Monaco et al. 2007; Starkenburg et al. 2009; Carlin et al. {\it in prep.}).

In the inner Galaxy, dynamical tracers such as tidal streams can provide strong constraints on the mass distribution of the baryonic components of the 
Milky Way (e.g., Font et al. 2001; Ibata et al. 2003; Koposov et al. 2009).
At  the larger radii $r \sim 15-60$ kpc traced by the orbit of the Sgr dwarf however  the gravitational potential is dominated by dark matter, and observations of the luminous tidal streams can
be used to constrain the shape, orientation, and mass of the Milky Way's dark halo (e.g., Ibata et al. 2001b;
Helmi 2004; Johnston et al. 2005; Law et al. 2005 [hereafter LJM05]; Fellhauer et al. 2006; Mart{\'{\i}}nez-Delgado et al. 2007).
The breadth and quality of the recent observational data is such that it has highlighted a 
`halo conundrum' (variously discussed by LJM05; Fellhauer et al. 2006; Mart{\'{\i}}nez-Delgado et al. 2007;  Newberg et al. 2007; and Yanny et al. 2009):
In a static axisymmetric Milky Way dark halo no one model has been capable of simultaneously reproducing 
both the angular position and distances/radial velocities of tidal debris in the Sgr leading arm.
Depending on which criterion a given study chose to focus, claims have been made in favor of 
an oblate halo (e.g., Johnston et al. 2005; Mart{\'{\i}}nez-Delgado et al. 2007),
an approximately spherical halo (e.g., Ibata et al. 2001b; Fellhauer et al. 2006), and a prolate halo (e.g., Helmi 2004).

As we demonstrated using massless test-particle orbits in Law, Majewski, \& Johnston (2009; hereafter LMJ09) however, 
it is possible to resolve this conundrum by adopting a triaxial halo in which the minor axis is approximately aligned with the line of sight to the Galactic Center.
In this contribution we extend the results of LMJ09 by performing detailed $N$-body simulations of the tidal disruption of the Sgr dwarf
within a range of Galactic halo parameterizations that were favored by test-particle orbits.
In \S \ref{basicapproach.sec} we outline our basic numerical approach, describing the generalized form of the Milky Way --- Sgr system along with an overview
of the observational constraints on the Sgr tidal streams.
We next describe our solution to the problem of constraining the basic orbital parameters of the Sgr stream (for a given Milky Way model) in \S \ref{sgrprop.sec}, 
using these various simulations to discriminate between different realizations of the underlying Galactic potential in \S \ref{mwprop.sec}.
The final parameters describing our best-fit model of the Milky Way --- Sgr system are summarized in \S \ref{overview.sec}.

In \S \ref{halodisc.sec} we discuss the best-fit model for the dark matter halo of the Milky Way that was derived in \S \ref{mwprop.sec}, and present possible theoretical
challenges to, alternatives for, and methods of testing the triaxial halo model.
We demonstrate 
in \S \ref{feh.sec}  that by adopting simple assumptions for the star formation history of Sgr it is possible to reproduce the trends in the MDF
observed along the stellar tidal streams.
We compare the best-fitting model of the Sgr streams to other observations of halo substructure that have previously been postulated to be associated with the Sgr stream
in \S \ref{otherobs.sec}, and conclude in \S \ref{summary.sec} with a summary of the successes and failures of our new model of the Milky Way --- Sgr system
and an overview of the key issues that need to be addressed in future work.

Throughout this paper
we adopt the heliocentric Sgr coordinate system ($\Lambda_{\odot}, B_{\odot}$) defined by Majewski et al. (2003) in which the 
longitudinal coordinate $\Lambda_{\odot} = 0^{\circ}$
in the direction of Sgr and increases along trailing tidal debris, and $B_{\odot}$ is positive towards the 
orbital pole $(l,b)_{\rm pole} \approx (274^{\circ}, -14^{\circ})$.


\section{Basic Approach}
\label{basicapproach.sec}

\subsection{Numerical Framework}
\label{model.sec}

We adopt a basic  formalism similar to that described by LJM05 and LMJ09.
The Milky Way is described by a smooth fixed gravitational potential
consisting of a Miyamoto-Nagai (1975) disk, Hernquist spheroid, and a logarithmic halo.
The gravitational potential is therefore specified directly through the isovelocity ellipsoids both for consistency with previous studies
and because these are more analytically accessible than working
with an underlying mass distribution from which the gravitational acceleration must be numerically derived.

The respective contribution of these components is given by:
\begin{equation}
        \Phi_{\rm disk}=- \alpha {GM_{\rm disk} \over
                 \sqrt{R^{2}+(a+\sqrt{z^{2}+b^{2}})^{2}}},
\label{diskeqn}
\end{equation}
\begin{equation}
        \Phi_{\rm sphere}=-{GM_{\rm sphere} \over r+c},
\label{bulgeqn}
\end{equation}
\begin{equation}
        \Phi_{\rm halo}=v_{\rm halo}^2 \ln (C_1 x^2 + C_2 y^2 +C_3 x y + (z/q_z)^2 + r_{\rm halo}^2)
\label{haloeqn}
\end{equation}
where $R/r$ are cylindrical/spherical radii respectively, and the various constants $C_1, C_2, C_3$ are given by
\begin{equation}
C_1 = \left(\frac{\textrm{cos}^2 \phi}{q_{1}^2} + \frac{\textrm{sin}^2 \phi}{q_{2}^2}\right)
\end{equation}
\begin{equation}
C_2 = \left(\frac{\textrm{cos}^2 \phi}{q_{2}^2} + \frac{\textrm{sin}^2 \phi}{q_{1}^2}\right)
\end{equation}
\begin{equation}
C_3 = 2 \, \textrm{sin}\phi \, \textrm{cos} \phi \left( \frac{1}{q_{1}^2} - \frac{1}{q_{2}^2}\right)
\end{equation}

As discussed by LMJ09, this form for the dark halo potential describes an ellipsoid rotated by an angle $\phi$ about the Galactic $Z$ axis, in
which $q_1$ and $q_2$ are the axial flattenings along the equatorial axes and $q_z$ is the axial flattening perpendicular to the Galactic disk.
We adopt the convention that $\phi = 0^{\circ}$ corresponds to $q_1/q_2$ coincident with the Galactic $X/Y$ axes respectively, 
and increases in the direction of positive Galactic longitude.
Since only the ratios between the various $q$ are important (as increasing/decreasing $q_1/q_2/q_z$ in tandem is largely degenerate with the total
halo mass normalization $v_{\rm halo}$), we fix $q_2 = 1.0$.

We held fixed any parameter that was found not to be significantly constrained by Sgr (i.e., that did not significantly change the path
of Sgr debris when varied).
Following Johnston et al. (1999) and LJM05, we take $M_{\rm disk}=1.0 \times
10^{11}$ $M_{\odot}$, $M_{\rm sphere}=3.4 \times 10^{10}$ $M_{\odot}$,
$a=6.5$ kpc, $b=0.26$ kpc, and $c=0.7$ kpc.  
In LJM05 we explored the possibility of scaling the total mass 
of the Galactic disk through the parameter $\alpha$ and found no compelling reason to adopt a value other than $\alpha = 1.0$.
We fix the distance to the Galactic Center to be  $R_{\odot} = 8$ kpc (e.g., Reid 1993; Groenewegen et al. 2008); in
LJM05 we explored the possibility of varying $R_{\odot}$, and found that for cases where the Sun lies approximately in the plane of the orbit
(as is the case for Sgr) any reasonable choice of $R_{\odot}$ does not have a particularly noticeable effect on the quality-of-fit of our simulations.
Additionally, Figure 4 of LJM05 suggests that the ideal choice of the scale length of the Galactic halo ($r_{\rm halo}$) is reasonably insensitive to the axial
flattening adopted, and that a single fixed value of $r_{\rm halo} = 12$ kpc should suffice.
The normalization of the dark halo mass via the scale parameter $v_{\rm halo}$ is therefore entirely specified (for a given choice of $q_1$, $q_2$, and $\phi$)
by the requirement that the speed of the Local Standard of Rest be $v_{\rm LSR} = 220$ \kms.
As for $R_{\odot}$, $v_{\rm LSR}$ is at present relatively unconstrained by observations of the Sgr stream (LJM05; although see discussion by 
Majewski et al. 2006; Carlin et al. {\it in prep})
and we simply fix $v_{\rm LSR}$ to this conventional value.
We note that if we were instead to adopt the proper motion of Sgr A$^{\ast}$ measured by Reid \& Brunthaler (2004) it would imply
$v_{\rm LSR} = 235.6$ \kms for our adopted value of $R_{\odot} = 8.0$ kpc.

In a recent contribution (LMJ09) we integrated massless test-particles representing the Sgr dwarf along orbits in the Galactic potential defined by Eqns. \ref{diskeqn} - \ref{haloeqn}
for a range\footnote{Note that LMJ09 only included the results from the range $q_1 = 1.0 - 1.8$, $q_z = 1.0 - 1.8$
in their Figures 3 and 4 since much of the $q_1 < 1.0$, $q_z < 1.0$ parameter space is degenerate with $q_1 > 1.0$, $q_z > 1.0$.  The non-degenerate parameter space with the minor axis oriented
along $q_z$ was strongly disfavored in the LMJ09 simulations.}
of parameters $\phi = 0 - 180^{\circ}$, $q_1 = 0.6 - 1.8$, $q_z = 0.6 - 1.8$.  While the accuracy of such simple orbit models is necessarily limited because 
actual tidal debris
from a massive satellite will deviate slightly from these orbits, LMJ09 were able to identify a region of parameter 
space
around $\phi \approx 90^{\circ}$, $q_1 \approx 1.5$, $q_z \approx 1.25$
generally favorable to producing a model satellite
capable of matching the observational constraints enumerated in \S \ref{obsdata.sec}.  
We note that while such test-particle orbits technically identify a second similarly acceptable region of parameter space 
around $\phi \approx 0^{\circ}$, $q_1 \approx 0.67$, $q_z \approx 0.83$, the physical axis ratios of this region 
are degenerate with the first: Both have the short axis approximately aligned with the Galactic $X$ axis, and
the longest with the Galactic $Y$.  We choose to focus on the $q_z > 1$ branch of the solution to avoid introducing degeneracy into our simulations.

Motivated by the successes of LMJ09, we perform $N$-body simulations in Galactic models with a range of parameters $\phi = 75^{\circ} - 115^{\circ}$, $q_1 = 1.1 - 1.6$, $q_z = 1.1 - 1.5$.
In each case, Sgr is integrated $\sim 8$ Gyr backwards in time along a simple test-particle orbit, and a $10^5$-particle model for the dwarf is integrated forward again in time
from this point to the present day.
The particles in our model of Sgr are initially distributed to generate a Plummer (1911) model
\begin{equation}
        \Phi_{\rm Sgr}=-{GM_{\rm Sgr, 0} \over \sqrt{r^2+r_{\rm 0}^2}},
\label{PlummerEqn}
\end{equation}
where $M_{\rm Sgr, 0}$ is the initial mass of Sgr and $r_{\rm 0} = (M_{\rm Sgr, 0}/10^9 M_{\odot})^{1/3}$ kpc is its scale length.
We have assumed spherical symmetry and isotropy of the orbits within the initial Sgr dwarf since: 1) There is no evidence yet of significant rotation in the main
body of the satellite, and 2) debris resulting from isotropic orbits will cover all phase-space regions accessible to non-isotropic satellites.
The simulation particles are not associated with specific dark or light components, but instead trace the total mass of the satellite.
The mutual interactions between these self-gravitating particles\footnote{The gravitational influence of the bound satellite is applied to
all $N$-body particles regardless of whether or not they are still bound to the dwarf.} in the external Milky Way potential are calculated using a
self-consistent field code (Hernquist \& Ostriker 1992).

\subsection{Overview of Observational Constraints}
\label{obsdata.sec}

We summarize below 
the observational constraints that we adopt on the physical properties of the Milky Way --- Sgr system, each of which is
discussed in detail in \S \ref{sgrprop.sec} - \ref{mwprop.sec}.
Generally, we have adopted these constraints because we consider them to be the strongest, most unambiguous available.
Note that we opt not to constrain the model to match the apparent bifurcation in the Sgr leading arm described by
Belokurov et al. (2006) since recent observational data presented by Yanny et al. (2009) suggest that feature may arise due
to substructure within the Sgr dwarf rather than the shape of the Galactic gravitational potential (see discussion in \S \ref{leadingprop.sec}
and \ref{bifurc.sec}).

While individual constraints are discussed in detail in \S \ref{sgrprop.sec}, we summarize below


\begin{enumerate}
\item Sgr must lie at the angular coordinates, distance, and radial velocity described in \S \ref{coreprop.sec}.
\item Sgr must have a velocity vector that lies within the orbital plane defined by recent trailing tidal debris.
\item The radial velocity dispersion among trailing tidal debris must match that observed by  Monaco et al. (2007) using high resolution echelle spectroscopy.
\item The run of radial velocities with orbital longitude along trailing tidal debris must match data presented by Majewski et al. (2004).
\item The run of radial velocities with orbital longitude along leading tidal debris must match data presented by Law et al. (2004; see also LJM05).
\item The angular location of leading tidal debris must match the high surface brightness stream observed by Belokurov et al. (2006) in the SDSS.
\item The angular width of leading tidal debris must match that observed by Belokurov et al. (2006).
\end{enumerate}

It turns out to be easiest to constrain the properties of the Sgr dwarf itself first (\S \ref{sgrprop.sec}), since many of them are largely independent of the exact form of the Galactic potential
(to within the range explored in this paper), and the rest may be constrained contingent upon a specific potential.  We then use
 these simulations to discriminate between different realizations of the underlying Galactic potential in \S \ref{mwprop.sec}.
We note  that  the only distance information explicitly incorporated in our fitting method is that used in setting the distance to the Sgr dwarf itself.
Our analysis therefore relies only upon the most
well-constrained observations of the tidal debris (namely, the angular coordinates and radial velocities) and is
insensitive to photometric
distance calibration uncertainties which have hindered confident interpretation of previous studies.  While we do not fit for the distances to Sgr tidal debris explicitly, 
we shall see in \S \ref{overview.sec} and \S \ref{mgiant_kmags.sec} that 
the final model is nonetheless an excellent match to the distances derived for the observational data using the most reasonable assumptions about distance calibrations.


\section{Constraining the Properties of the Sgr dSph}
\label{sgrprop.sec}

\subsection{Core Properties}
\label{coreprop.sec}

We assume that the Sgr dwarf is located at Galactic coordinates $(l,b) = (5.6^{\circ}, -14.2^{\circ})$ (Majewski et al. 2003).
Estimates of the distance to Sgr typically range from $24-28$ kpc (see Table 2 of Kunder \& Chaboyer 2009 for a summary);
we adopt the distance scale $D_{\rm Sgr} = 28$ kpc  from Siegel et al. (2007) based on {\it Hubble Space Telescope} ({\it HST})/ACS
 analysis of the stellar populations in the Sgr core. 
In LJM05 (which adopted $D_{\rm Sgr} = 24$ kpc) we explored the effect of varying $D_{\rm Sgr}$ and found it to be largely
degenerate with other parameters (such as $R_{\odot}$) and not particularly well-constrained, although models generally favored
higher values of $D_{\rm Sgr}/R_{\odot}$.  We therefore simply fix $D_{\rm Sgr} = 28$ kpc 
to minimize the number of free parameters in our model.
This places the Sgr dwarf at Galactic coordinates\footnote{We 
adopt a right-handed Galactic Cartesian coordinate system with origin at the Galactic Center.  All velocities are given
in the Galactic Standard of Rest (GSR) frame, with respect to which the Sun has a peculiar motion $(U,V,W) = (9, 12+220, 7)$ \kms.}
$(X, Y, Z) = (19.0, 2.7, -6.9)$ kpc.
The radial velocity of Sgr is taken to be $v_{\rm rad} = 171$ \kms (Ibata et al. 1997), and the dwarf is assumed
to be presently moving toward the Galactic plane (Irwin et al. 1996; Dinescu et al. 2005).

\subsection{Constraining the Orbital Plane}
\label{orbitalpole.sec}

With the advantages of an internally homogeneous selection function, M giants
drawn from the 2MASS survey clearly define the orbital plane of the Sgr stream,
especially those high latitude M giants free of thick disk contamination 
and spanning $\sim 150^{\circ}$ across the South Galactic Hemisphere (Majewski et al. 2003).
This trailing stream is visible as it wraps from the Sgr core
($\Lambda_{\odot} = 0^{\circ}$) through the South Galactic Hemisphere towards the
Galactic anticenter ($\Lambda_{\odot} \sim 150^{\circ}$) --- near the position of the 
previous apocenter of the Sgr orbit ($\Lambda_{\odot} \sim 170^{\circ}$) .
The stream appears to dwindle rapidly at $\Lambda_{\odot} \gtrsim 140^{\circ}$ however, largely because the increasing age and 
decreasing metallicity (e.g., Chou et al. 2007, 2009) of this older tidal debris
results in a decline in the prevalence of the M-giant tracer population.

\begin{figure*}
\plotone{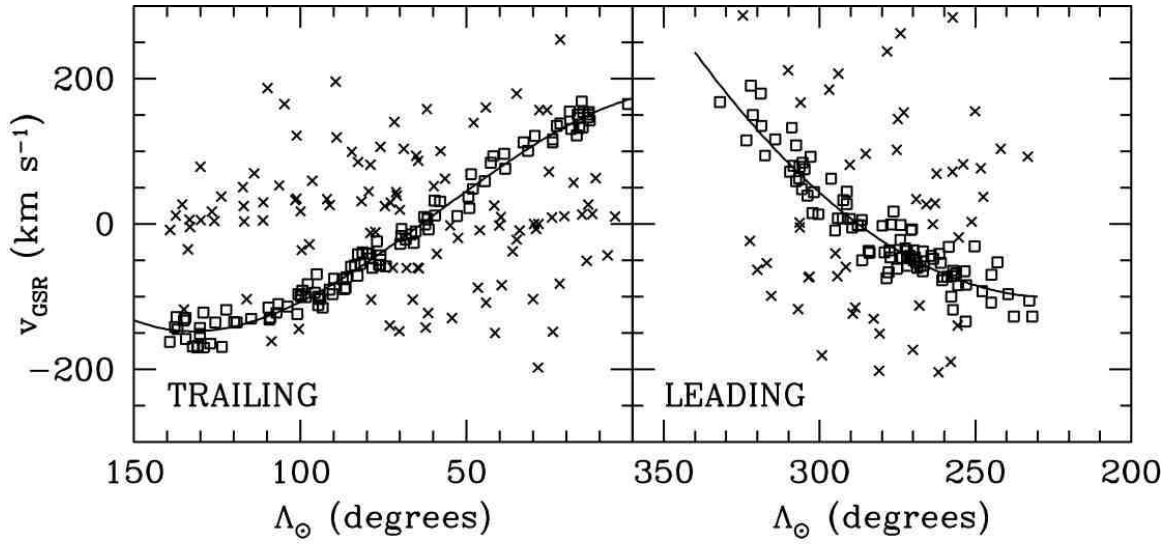}
\caption{Radial velocities for candidate Sgr stream stars in the leading (right-hand panel) and trailing (left-hand panel) tidal streams from Law et al. (2004, 2005)
and Majewski et al. (2004) respectively as a function of Sgr orbital longitude $\Lambda_{\odot}$.  Open squares (crosses) represent stars meeting (not meeting)
our iterative $2.5\sigma$ rejection algorithm, the solid lines represent the converged fourth-order polynomial fits to the velocity trends.}
\label{rvcull.fig}
\end{figure*}

The trend of radial velocities along the trailing stream is well-defined (Majewski et al. 2004), enabling us to use kinematical information to prune Galactic interlopers and
select a sample of extremely high-likelihood Sgr trailing arm stars.
We draw such a sample from the Majewski et al. (2004)  catalog, including all stars in the 
range $\Lambda_{\odot} = 0^{\circ} - 140^{\circ}$ for which radial velocity data has been obtained.\footnote{The sampling
criteria of Majewski et al. (2004) required target stars to have infrared color $J-K_s = 1.0 - 1.1$, and be located within 5 kpc of the Sgr plane defined by Majewski et al. (2003).}
We fit a fourth-order polynomial to the radial velocity data as a function of orbital longitude, using an iterative $2.5\sigma$ rejection algorithm to produce our final clean sample
of trailing arm stars (Figure \ref{rvcull.fig}, left panel).\footnote{We provide an online-only table of the coordinates and radial velocities in this clean sample at
http://www.astro.virginia.edu/$\sim$srm4n/Sgr.}
Note that we have
not applied an explicit distance selection cut to the sample of stars plotted in Figure \ref{rvcull.fig}; if we were to do so it would remove a majority of the crosses from the left-hand panel
as these stars lie significantly closer than the Sgr trailing arm stream (see, e.g., Figure 2 of Majewski et al. 2004).

The visible and well-defined portion of the trailing arm seen in the Southern Galactic Hemisphere 2MASS M giants is particularly
 important because it is dynamically young (torn off in the last $\sim$ 0-2 pericentric passages, depending on the assumed mass of Sgr)
and has therefore  experienced very little differential precession from the main body of Sgr (e.g., Helmi et al. 2004; Johnston et al. 2005).
Observations (Majewski et al. 2003) indicate that the trailing debris stream is well-confined to a single plane, and numerical simulations show that $N$-body debris in the range
$\Lambda_{\odot} = 0^{\circ} - 140^{\circ}$ have angular momentum
vectors aligned with that of the present satellite to within 1 degree  for a wide range of models of the Galactic gravitational potential (see, e.g., Figures 4 and 5 of Johnston et al. 2005).

The lack of orbital precession in the trailing stream allows us to determine robustly the instantaneous angular momentum vector of Sgr.
Using the sample of trailing arm stars defined above, we use a GC3 cell count technique (Johnston et al. 1996) 
to define the plane that best fits all of the trailing arm stars and the present position of the Sgr dwarf.
For any given orbital pole $(l_{\rm p}, b_{\rm p})$ the distance $D_i$ of star $i$ from the plane defined by the pole is
\begin{equation}
D_i = \hat{n} \cdot (\vec{r}_i - \vec{r}_{\rm Sgr})
\end{equation}
where $\hat{n}=(\textnormal{cos}\,b_{\rm p} \, \textnormal{cos} \,l_{\rm p}, \textnormal{cos}\,b_{\rm p} \, \textnormal{sin}\,l_{\rm p}, \textnormal{sin}\,b_{\rm p})$
is the orbital pole vector,
$\vec{r}_i$ is the vector from the Galactic Center to star $i$, and $\vec{r}_{\rm Sgr}$ is the vector from the Galactic Center to Sgr.

\begin{figure}
\plotone{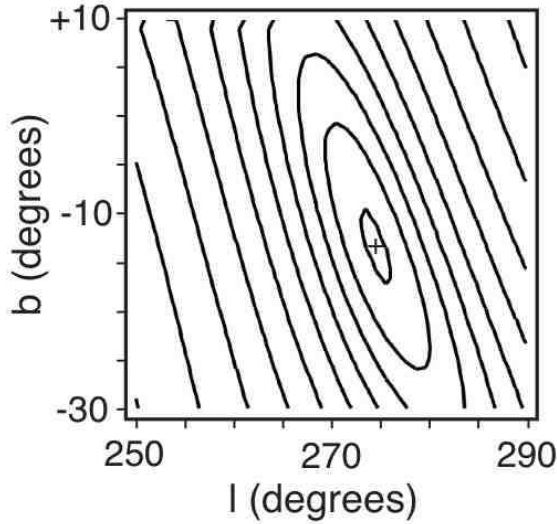}
\caption{GC3 orbital pole plot showing the total value of $D$ for a grid of orbital poles.  Contours are logarithmic in $D$ and indicate
the best-fitting orbital pole lies at $(l_{\rm p}, b_{\rm p}) = (273.8^{\circ}, -14.5^{\circ})$.}
\label{pole.fig}
\end{figure}

The best fitting plane  is that which minimizes the sum $D^2 = \Sigma_i D_i^2$
over the 108 data points in the clean sample of trailing arm stars shown in Figure \ref{rvcull.fig}.
In Figure \ref{pole.fig} we plot $D$ as a function of $(l_{\rm p}, b_{\rm p})$, and demonstrate that the best-fitting orbital pole
is $(l_{\rm p}, b_{\rm p}) = (273.8^{\circ}, -14.5^{\circ})$.  Unsurprisingly (since both the sample of stars and the technique
were very similar), this is nearly identical to the trailing arm pole calculated by Johnston et al. (2005), although it differs slightly from the pole adopted by LJM05, who
used the pole derived by Majewski et al. (2003) from fitting to both leading and trailing Sgr tidal debris.

\subsection{Constraining the Orbital Speed of Sgr}
\label{propermotion.sec}

Since the  radial velocity vector $\vec{v}_{\rm rad}$ and orbital pole (\S \ref{orbitalpole.sec}) of Sgr are known, the 3D space velocity of 
Sgr ($\vec{v}_{\rm Sgr} = \vec{v}_{\rm rad} + \vec{v}_{\rm tan}$, where by definition $\vec{v}_{\rm rad} \cdot \vec{v}_{\rm tan} = 0$) can be
uniquely specified by the magnitude $v_{\rm tan}$ of the velocity of Sgr tangential to the line of sight:
\begin{equation}
\vec{v}_{\rm Sgr} \cdot \hat{n} = \vec{v}_{\rm rad} \cdot \hat{n} + \vec{v}_{\rm tan} \cdot \hat{n} = 0
\end{equation}

In particular, for each choice of the Galactic halo parameters $\phi, q_1, q_z$ there is a value of $v_{\rm tan}$ that produces a trailing tidal stream whose velocities best match those
shown in Figure \ref{rvcull.fig}.  We converge upon this value in a manner similar to LJM05 by 
minimizing the $\chi^2$ statistic:
\begin{equation}
\chi_{v,{\rm trail}}^2 =  \frac{1}{n_{v,{\rm trail}} - 3} \sum_i \frac{(v_{\rm model} [i] - v_{\rm obs,trail} [i])^2}{\sigma_{v}^2}
\label{chi_vtrail.eqn}
\end{equation}
where $v_{\rm model}[i]$ is determined via linear interpolation of the model to the $\Lambda_{\odot}[i]$ of each 
of the $n_{v,{\rm trail}} =108$
velocity data points $v_{\rm obs, trail}[i]$.  

There are two components governing the apparent velocity width of the stream; physical dispersion and observational uncertainty in the derived radial velocities.
Majewski et al. (2004) find a net dispersion of 12 \kms, composed of  6 \kms intrinsic uncertainty in the velocity measurements  and an underlying
10 \kms physical dispersion in the stream.  For the purposes of Eqn. \ref{chi_vtrail.eqn} it is irrelevant what the source of the dispersion is 
(since both will have identical effects governing the offset of stars from the orbital path) and we take $\sigma_v = 12$ \kms.
Rather than perform computationally intensive $N$-body simulations to constrain $v_{\rm tan}$ for every choice of $\phi, q_1, q_z$, we instead use test-particle orbits
(see, e.g., LMJ09)
to estimate the necessary value.  Based on $\sim 150$ $N$-body simulations with a wide range of Sgr masses and forms for the Galactic potential, 
we find a typical correction factor where the best-fit $v_{\rm tan}$ for $N$-body models is systematically 4 \kms slower than 
that implied by $\chi^2$ minimization using such test-particle orbits.

\begin{figure}
\epsscale{1}
\plotone{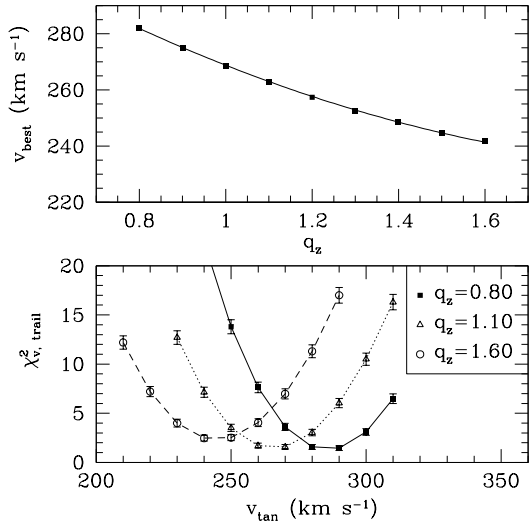}
\caption{\scriptsize Bottom panel: Reduced $\chi_{v,{\rm trail}}^2$ statistic for the fit of a test-particle orbit to trailing arm velocity data as a function of the
Sgr velocity $v_{\rm tan}$ tangential to the line of sight for fixed $\phi = 90^{\circ}$, $q_1 = 1.0$ and three choices of $q_z = 0.80/1.10/1.60$.  Solid/dotted/dashed lines respectively indicate third-order
polynomial fits to the discrete data points.
Top panel: Best-fit Sgr tangential velocity $v_{\rm best}$ (defined by minimizing $\chi_{v,{\rm trail}}^2$) as a function of $q_z$ for $\phi = 90^{\circ}$, $q_1 = 1.0$.
The solid line represents a polynomial fit to the discrete measurements.}
\label{vtan.fig}
\end{figure}

Considering the example case where $\phi = 90^{\circ}$ and $q_1 = 1.0$,
in Figure \ref{vtan.fig} (bottom panel) we illustrate the dependence of $\chi_{\rm v, trail}^2$ on $v_{\rm tan}$ for three different values of $q_z$.  
Note that there is a well-defined
best value (which we call $v_{\rm best}$) of  $v_{\rm tan}$ for each $q_z$; we estimate the location of the minimum by fitting a simple third-order polynomial to $\chi_{\rm v, trail}^2$ 
as a function of $v_{\rm tan}$.  The resulting values of $v_{\rm best}$ are well-behaved as a function of $q_z$, as illustrated in the top panel of Figure \ref{vtan.fig}.

For each choice of $\phi$ and $q_1$ it is possible to construct a similar relation.
For $\phi = 90^{\circ}$ the relation between $v_{\rm best}$ and $q_z$ is practically unchanged (to within $\sim 1 - 2$ \kms) for any choice of $q_1 = 0.8 - 1.6$.
For other choices of $\phi$ and $q_1$ the deviation of the relation from that shown in Figure \ref{vtan.fig} grows as $\phi$ departs from $90^{\circ}$, but for the range of $\phi = 75 - 115^{\circ}$
considered here these relations differ by less than 10 \kms.


\subsection{Constraining the Mass of Sgr}

Data in the longitude range 
$\Lambda_{\odot} = 25^{\circ} - 90^{\circ}$ consist of a relatively orderly sample of trailing arm
stars as they pass through the south Galactic cap moving predominantly transverse to our line of sight, and stars in this range can provide the most accurate
measurement of the intrinsic velocity dispersion of the stream\footnote{There is a slight thickening in the distribution of radial velocities at $\Lambda_{\odot} \gtrsim 100^{\circ}$
because of the angular superposition of tidal debris from different epochs (i.e., stripped from Sgr on different pericentric passages) that occurs when this debris
`piles up' near the previous apogalactic point along the orbit.}.  Majewski et al. (2004) performed such an analysis, concluding that for a reasonable estimate of the observational
uncertainty that the intrinsic velocity dispersion of the stream is $\sigma_{v,{\rm obs}} = 10.0 \pm 1.7$ \kms.  More recent, higher spectral resolution observations\footnote{Note that while 
we use the Monaco et al. (2007) velocity dispersion to derive the mass of Sgr, we used
the Majewski et al. (2004) velocity dispersion in Equation \ref{chi_vtrail.eqn} because this was the value appropriate to that more extensive sample of stars used to define
the velocity trend of the trailing arm.} by Monaco et al. (2007)
have revised this value slightly to $\sigma_{v,{\rm obs}} = 8.3 \pm 0.9$ \kms.

\begin{figure}
\plotone{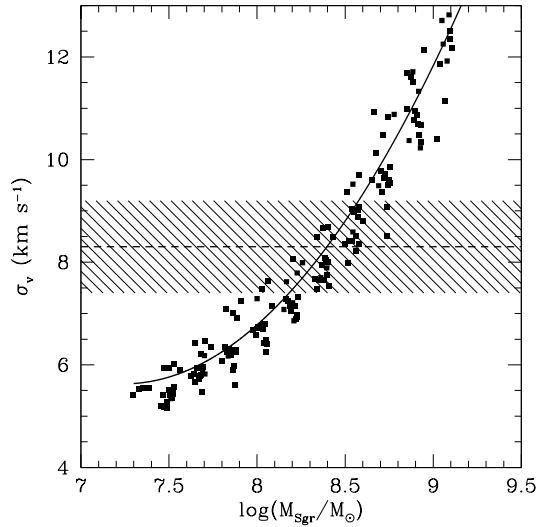}
\caption{\scriptsize Velocity dispersion $\sigma_{\rm v}$ along the Sgr trailing arm (filled boxes) as a function of the final bound Sgr mass $M_{\rm Sgr}$ 
for $\sim 160$ $N$-body simulations with different realizations of the Galactic potential, Sgr initial mass $M_{\rm Sgr, 0}$, and interaction time.
The hatched area and dashed line indicate the observational range $\sigma_{\rm v} = 8.3 \pm 0.9$ \kms measured from high-resolution
spectra by Monaco et al. (2007).  The solid curve represents a polynomial fit to the simulated data.}
\label{massfig.fig}
\end{figure}

Since the velocity dispersion in the tidal debris streams reflects the mass present within the tidal radius of Sgr on the orbit immediately prior to that debris
becoming unbound (see, e.g., Johnston et al. 1996; Johnston 1998; Pe{\~n}arrubia et al. 2008), it 
can therefore be used to constrain the mass of the dSph (e.g., LJM05).
We performed a series of $\sim$ 160 $N$-body simulations in which satellites with 10 different initial masses ranging from 
$M_{\rm Sgr, 0} = 6.67 \times 10^7 - 2.57 \times 10^9 M_{\odot}$
were integrated along reasonable orbits for various lengths of time in different realizations of the Milky Way gravitational potential.
The resulting relation between the final bound Sgr mass $M_{\rm Sgr}$ and the velocity dispersion $\sigma_{\rm v}$ is shown in Figure \ref{massfig.fig}.
Extrapolating from the best-fit
polynomial to the curve, we conclude that a present bound  mass $M_{\rm Sgr} = 2.5^{+1.3}_{-1.0} \times 10^8 M_{\odot}$
produces the best fit to the observed velocity dispersion.  Adopting a typical mass-loss history (generally similar along reasonable orbits in the range
of Galactic potentials explored here), this implies that the original mass of Sgr was 
$M_{\rm Sgr, 0} = 6.4^{+3.6}_{-2.4} \times 10^8 M_{\odot}$ if it has been orbiting in the Galactic potential for $\sim 8$ Gyr.  We adopt
$M_{\rm Sgr, 0} = 6.4 \times 10^8 M_{\odot}$, corresponding to an initial Sgr scale length 
$r_{\rm 0} = 0.85$ kpc (see Equation \ref{PlummerEqn}).

\section{Constraining the Properties of the Milky Way Halo}
\label{mwprop.sec}

We have now constrained the properties of Sgr itself as  functions of the Galactic parameters $\phi, q_1, q_z$.  In the following subsections we go on to explore
within which of these potentials the observational characteristics of the tidal streams are best reproduced.  

\subsection{Sgr Trailing Arm Properties}

As defined in \S \ref{propermotion.sec}, 
we use the statistic $\chi_{\rm v, trail}$ to quantify the fit of a given $N$-body simulation to the trailing arm velocity trend.  While $\chi_{\rm v, trail}$ has already been minimized
for a given choice of $\phi, q_1, q_z$ by determining the orbital velocity of the satellite, the precise minimum value can vary slightly between different halo realizations
and we therefore also use it as a constraint on the properties of the Milky Way.

\subsection{Sgr Leading Arm Properties}
\label{leadingprop.sec}

Although both the leading and trailing tidal tails are traced by the M-giant population, the location of Sgr slightly past pericenter along its orbit
in combination with foreshortening of the 
orbital rosettes caused by our location 8 kpc from the center of the Milky Way means that 
while the trailing arm extends for $\sim 150^{\circ}$ across the southern Galactic hemisphere  the leading arm 
is clearly visible only for $\sim 75^{\circ}$ (a significant section of which is obscured  as it passes behind the Galactic disk/bulge).
The exact track of leading Sgr debris as it passes through the north Galactic cap towards the anticenter is therefore uncertain in the 2MASS data since the M-giant tracer population
is no longer as prominent in this older section of the stream.
In contrast, SDSS imaging is sensitive to older Sgr populations, and clearly shows (Belokurov et al. 2006; Newberg et al. 2007; Cole et al. 2008) the Sgr leading arm
arcing through the northern Galactic cap towards re-entry into the Galactic disk in the direction of the anticenter.

Belokurov et al. (2006) initially suggested that there may be a bifurcation in the angular position of the Sgr stream 
in the northern Galactic hemisphere, with a high surface brightness `A' stream representing the primary wrap of the leading tidal arm 
passing through the coordinates $(\alpha,\delta) \sim (160^{\circ}, 20^{\circ})$ and a lower surface-brightness `B' stream  representing a secondary wrap
of the trailing tidal arm offset to higher declination around $(\alpha,\delta) \sim (160^{\circ}, 30^{\circ})$ (see also Fellhauer et al. 2006).
We adopt the positions of the `A'-stream survey fields
of Belokurov et al. (2006)  along this high-surface brightness 
branch as our constraint on the orbital path of leading Sgr tidal debris.  These 17 pointings are tabulated in Table \ref{SDSS.tab}.
Note that Belokurov et al. (2006) adopted a photometric
distance scale calibrated against a  distance to Sgr of 25 kpc.  The distances listed in Table \ref{SDSS.tab} have been rescaled to our adopted
$D_{\rm Sgr} = 28$ kpc by multiplying by a factor of 1.12.

We note that the dynamical origin of the lower surface-brightness B stream is uncertain: Recent observational data presented by Yanny et al. (2009)
indicate that the A and B streams have indistinguishable trends in distance, radial velocity, metallicity, and mix of stellar types.  
This suggests that the A and B branches represent debris stripped from Sgr at similar times, and that the bifurcation may likely be
within a single phase of tidal debris, originating in 
substructure
within the initial Sgr dwarf (see discussion in \S \ref{bifurc.sec}).  Since we do not attempt to constrain the internal structure of the Sgr dwarf (we focus here on exploring the effects
of triaxiality in the gravitational potential of the Milky Way, and such an effort is beyond
the scope of this contribution), we therefore do not explicitly constrain our model to match the B branch of the bifurcation.

Numerically, we define the statistic
\begin{equation}
\chi_{\delta}^2 = \frac{1}{n_{\delta} - 3} \sum_i \frac{(\delta_{\rm model}[i] - \delta_{\rm obs}[i])^2}{\sigma_{\delta}^2}
\end{equation}
describing the accuracy with which the declination of the $N$-body model traces the A stream of Belokurov et al. (2006), and
\begin{equation}
\chi_{w}^2 = \frac{(w_{\delta, {\rm model}} - w_{\delta, {\rm obs}})^2}{\sigma_{w}^2}
\end{equation}
describing the accuracy with which the angular width of the $N$-body model matches the SDSS data at a right ascension $\alpha = 180^{\circ}$
(the non-spherical gravitational potential gives rise to  angular broadening of the  leading debris stream in this region due to differential precession).
Based on the data presented by Belokurov et al. (2006), we estimate that the $1\sigma$ angular width of the stream as seen in SDSS is $w_{\delta, {\rm obs}} = 5.2^{\circ}$,
with an uncertainty $\sigma_w = 1.3^{\circ}$.
In contrast to Equation \ref{chi_vtrail.eqn} (which used $\Lambda_{\odot} [i]$ as the variable coordinate), $\delta_{\rm model}[i]$ is determined via linear interpolation of the model to the 
right ascension $\alpha[i]$ of each of the
$n_{\delta} = 17$ SDSS survey fields along the main branch of the Sgr stream, each of which is assumed to have an angular uncertainty.\footnote{This 
is a rough estimate 
of the uncertainty in the centroid of the stream at a given right ascension based on the apparent width of the stream.}
 $\sigma_{\delta} = 1.9^{\circ}$.

While the trend of leading tidal debris is difficult to trace with 2MASS imaging data alone, the stream can be recovered using spectroscopy to identify the radial velocity
signature of the stream.  In Figure \ref{rvcull.fig} we show our selection function for stars in the Sgr leading arm 
with $K_s > 9$ (see \S \ref{mgiant_kmags.sec})
from the observational data presented by Law et al. (2004; see also LJM05), 
using a method similar to that adopted for trailing arm stars.  This velocity trend has recently been confirmed by 
spectroscopic observation of a larger
number of stars drawn from the SDSS imaging of the leading tidal stream (Yanny et al. 2009), in contrast with the preliminary conjecture by Newberg et al. (2007)
that the velocity trend may be artificially produced by confusion in the leading arm M-giant sample.
This gives our final constraint on the stream
\begin{equation}
\chi_{v,{\rm lead}}^2 = \frac{1}{n_{v,{\rm lead}} - 3} \sum_i \frac{(v_{\rm model} [i] - v_{\rm obs,lead} [i])^2}{\sigma_{v}^2}
\end{equation}
where  $v_{\rm model}[i]$ is determined via linear interpolation of the model to the orbital longitude
$\Lambda_{\odot}[i]$ of each of the  $n_{v,{\rm lead}} = 94$  observational data points in the 2MASS leading arm radial velocity sample.
As for the trailing arm velocity constraint, we adopt an uncertainty $\sigma_v = 12$ \kms for each radial velocity measurement  representing a combination of observational uncertainty and intrinsic stream width.

\begin{deluxetable}{lcc}
\tablecolumns{3}
\tablewidth{0pc}
\tabletypesize{\scriptsize}
\tablecaption{Leading Arm SDSS Constraints\tablenotemark{a}}
\tablehead{
\colhead{RA} & \colhead{Dec} & \colhead{Distance\tablenotemark{b}}\\
\colhead{(degrees)} & \colhead{(degrees)} & \colhead{(kpc)}}
\startdata
215 & 4 & $51.9^{+14.6}_{-2.2}$\\
210 & 4 & $51.0^{+15.5}_{-2.8}$\\
205 & 4 & $50.0^{+6.0}_{-4.5}$\\
200 & 4 & $47.3^{+3.9}_{-2.7}$\\
195 & 7 & $43.7^{+2.9}_{-1.7}$\\
190 & 9.5 & $41.1^{+5.5}_{-4.2}$\\
185 & 11.2 & $38.2^{+3.9}_{-2.5}$\\
180 & 13 & $35.7^{+2.7}_{-1.7}$\\
175 & 13.75 & $32.5^{+3.6}_{-2.7}$\\
170 & 15 & $30.3^{+2.5}_{-1.8}$\\
165 & 16 & $29.4^{+1.9}_{-1.1}$\\
150 & 18.4 & $22.6^{+2.3}_{-1.8}$\\
145 & 19 & $21.9^{+3.5}_{-2.6}$\\
140 & 19.4 & $20.5^{+1.8}_{-1.2}$\\
135 & 19.5 & ...\\
130 & 19.5 & $18.8^{+3.0}_{-2.3}$\\
125 & 19.5 & ...\\
\enddata
\tablenotetext{a}{Survey fields are from Belokurov et al. (2006).}
\tablenotetext{b}{Distances and uncertainties are taken from Belokurov et al. (2006) and scaled to reflect our choice of distance modulus to the Sgr core.}
\label{SDSS.tab}
\end{deluxetable}


\subsection{Combined Constraints}
\label{combinedconst.sec}

In general, the parameters $\phi, q_1, q_z$ are not separable and must be constrained simultaneously by 
$\chi_{\rm v, trail}$, $\chi_{\rm v, lead}$, $\chi_{\delta}$, and $\chi_w$.
We combine these four constraints into a single statistic:
\begin{equation}
\chi^2 = \chi_{\rm v, trail}^2 + \chi_{\rm v, lead}^2 + \chi_{\delta}^2 + \chi_w^2
\end{equation}
and seek the combination of $\phi, q_1, q_z$ that gives the global minimum for this combined statistic.

\begin{figure*}\epsscale{1}
\plotone{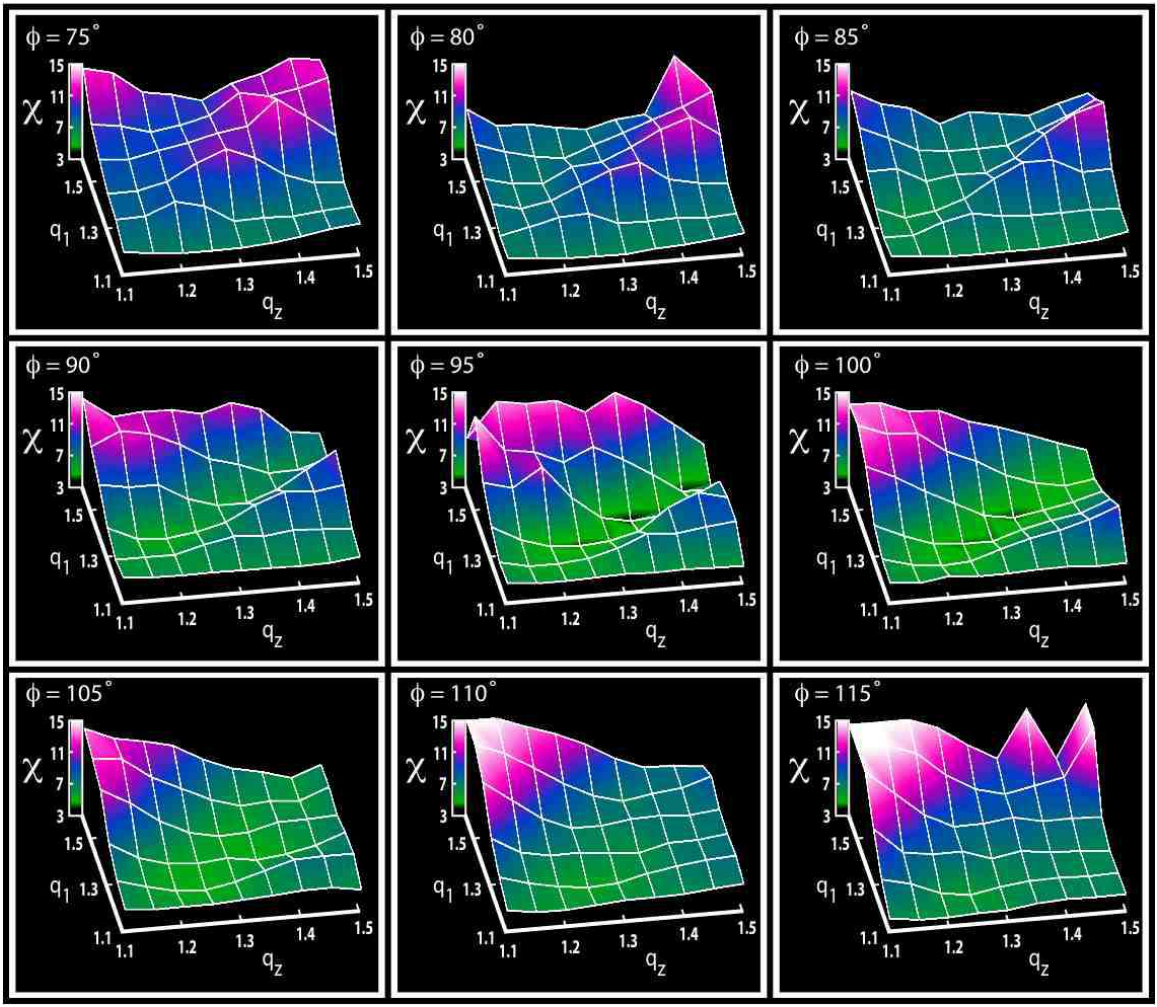}
\caption{The combined quality-of-fit criterion $\chi$ is plotted as a function of the dark halo axial scales $q_1$ and $q_z$ for nine values of the rotation angle 
in the range $\phi = 75^{\circ} - 115^{\circ}$.
The vertical stretch varies between panels, but the color map is identical.  The best fit is achieved for dark green/black regions around $q_1 \approx q_z$ in the central panels
$\phi = 95^{\circ}$ and $100^{\circ}$.
The range of $\chi$ shown on these plots belie the large range in quality of fit of the simulations:
While $\chi \lesssim 4$ (bright green --- black regions) represents a reasonable fit to all of the observational 
criteria, $\chi \gtrsim 10$ (blue and magenta regions) are extremely discrepant with at least one criterion.}
\label{GOF.fig}
\end{figure*}

We performed a series of $\sim 500$ $N$-body simulations covering the range $\phi = 75^{\circ} - 115^{\circ}$,  $q_1 = 1.1 - 1.6$, $q_z = 1.1 - 1.5$
with a grid spacing of $5^{\circ}$ in $\phi$, 0.05 in $q_z$, and 0.1 in $q_1$.
The resulting values of $\chi$ as a function of $\phi, q_1, q_z$ are shown in Figure \ref{GOF.fig} and demonstrate a general
preference for values of $q_1 \approx q_z$ in the range $\phi \approx 90^{\circ} - 105^{\circ}$.    
Interpolating along the grid surface, the optimal
fit ($\chi = 3.4$) is achieved for $\phi = 97^{\circ}$, $q_1 = 1.38$, $q_z = 1.36$.
We use these parameters for our final best-fit $N$-body simulation.

Since this optimal fit has $\chi > 1$ there are obviously some residual imperfections in the model because different observational constraints favor different
regions of parameter space.  The greatest mismatch lies in the leading arm radial velocities ($\chi_{v,{\rm lead}} = 2.7$), followed by the leading arm
angular coordinates ($\chi_{\delta} = 1.6$),  the trailing arm radial velocities ($\chi_{v,{\rm trail}} = 1.3$), and lastly the angular width of the leading
stream ($\chi_{w} = 0.9$).  For comparison, the spherical halo model discussed by LJM05 is clearly a poor fit to the observational data 
(see Figure 10 of LJM05) and has $\chi \approx 9$.
In general we deem simulations with $\chi \gtrsim 5$ to be visibly poor fits. 

The optimal shape for the Galactic dark halo therefore appears to be an approximately oblate ellipsoid whose short axis is located in the Galactic disk
along the direction $(l,b) = (7^{\circ},0^{\circ})$.  We note that this differs somewhat from that derived by LMJ09, who found a more strongly triaxial ellipsoid
with $\phi = 90^{\circ}$, $q_1 = 1.50$, $q_z = 1.25$ by matching observed tidal debris to the orbital path of 
a point mass orbiting in the Galactic potential.
It is well known, however,  that actual leading/trailing arm debris
falls inside/outside of the orbit respectively (e.g., Eyre \& Binney 2009), meaning that
a slightly different shape will be derived for the Galactic halo when attempting to match simulated debris to observed debris (as done in this contribution), rather
than a simulated orbit to observed debris (as in LMJ09).

\section{Model Overview}
\label{overview.sec}

\subsection{Model Summary}

\begin{figure*}
\epsscale{0.8}
\plotone{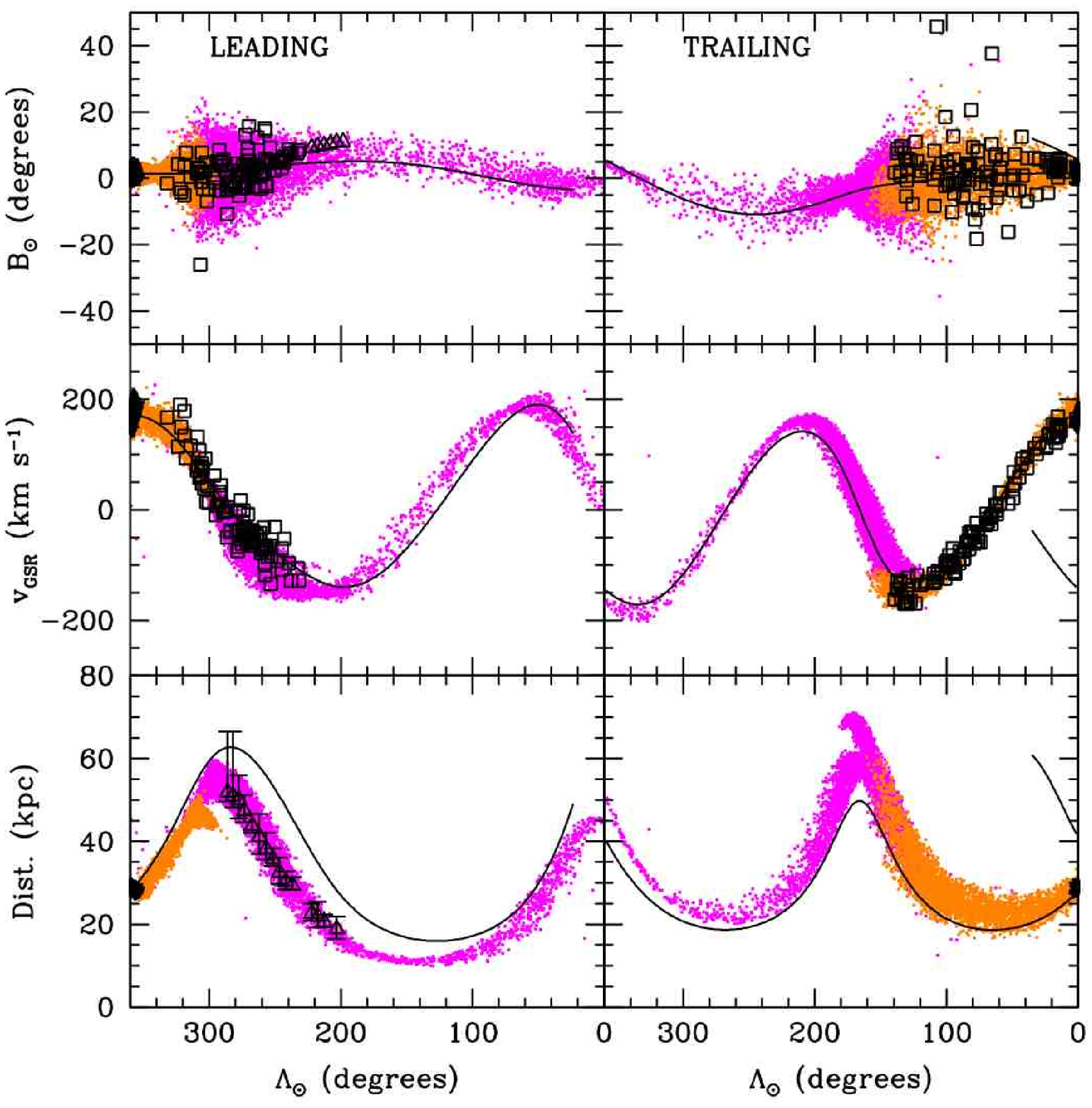}
\caption{\scriptsize The best fit $N$-body debris model (colored points) in a triaxial halo, overplotted with observational constraints from 2MASS + SDSS (open symbols). 
The left/right  columns plot leading/trailing arm debris respectively.
Colors represent the time at which a given debris particle became unbound from Sgr; for clarity we 
plot only debris from the most recent $\sim$ 3 Gyr of tidal disruption (see color-coding key in Figure \ref{OrbitPlot.fig} for details)
which best correspond to the observational data shown.
Open triangles represent SDSS A-stream data, open squares leading/trailing arm 2MASS M-giants culled from Figure \ref{pole.fig}.
The solid line represents the orbital path of the Sgr dSph core 1.2 Gyr ahead
and behind its current position.
The Sgr dwarf is currently located at orbital longitude/latitude $(\Lambda_{\odot}, B_{\odot}) = (0^{\circ}, 0^{\circ})$.
}
\label{ModelSummary.fig}
\end{figure*}

In LJM05, we presented possible models for the tidal disruption of Sgr under three different assumptions; that the Galactic halo was axisymmetric
in the disk plane and either oblate, spherical, or prolate in the direction perpendicular to the disk.
Neither of these three models were entirely satisfactory as they could not simultaneously reproduce the angular position, distance, and radial velocity trends of the leading arm.
In contrast, our new model,\footnote{A complete data file for the dynamical and chemical characteristics of Sgr debris from this best-fit model is provided
on the Web at http://www.astro.virginia.edu/$\sim$srm4n/Sgr.} orbiting in a {\it non-axisymmetric} dark matter halo, reproduces all of these constraints to reasonable accuracy
as demonstrated by Figure \ref{ModelSummary.fig}.
Indeed, we note that even though we didn't explicitly fit to the distances of the SDSS leading arm data from Belokurov et al. (2006), the new model nonetheless matches them to within the uncertainty
of the photometric calibration.
While it is not possible to test the agreement of this new model with the distances of the M-giant sample directly (due to the evolution of the MDF along the Sgr tidal streams), 
we demonstrate in \S \ref{mgiant_kmags.sec} that the model satisfactorily reproduces their apparent magnitude trend.

\begin{figure}
\plotone{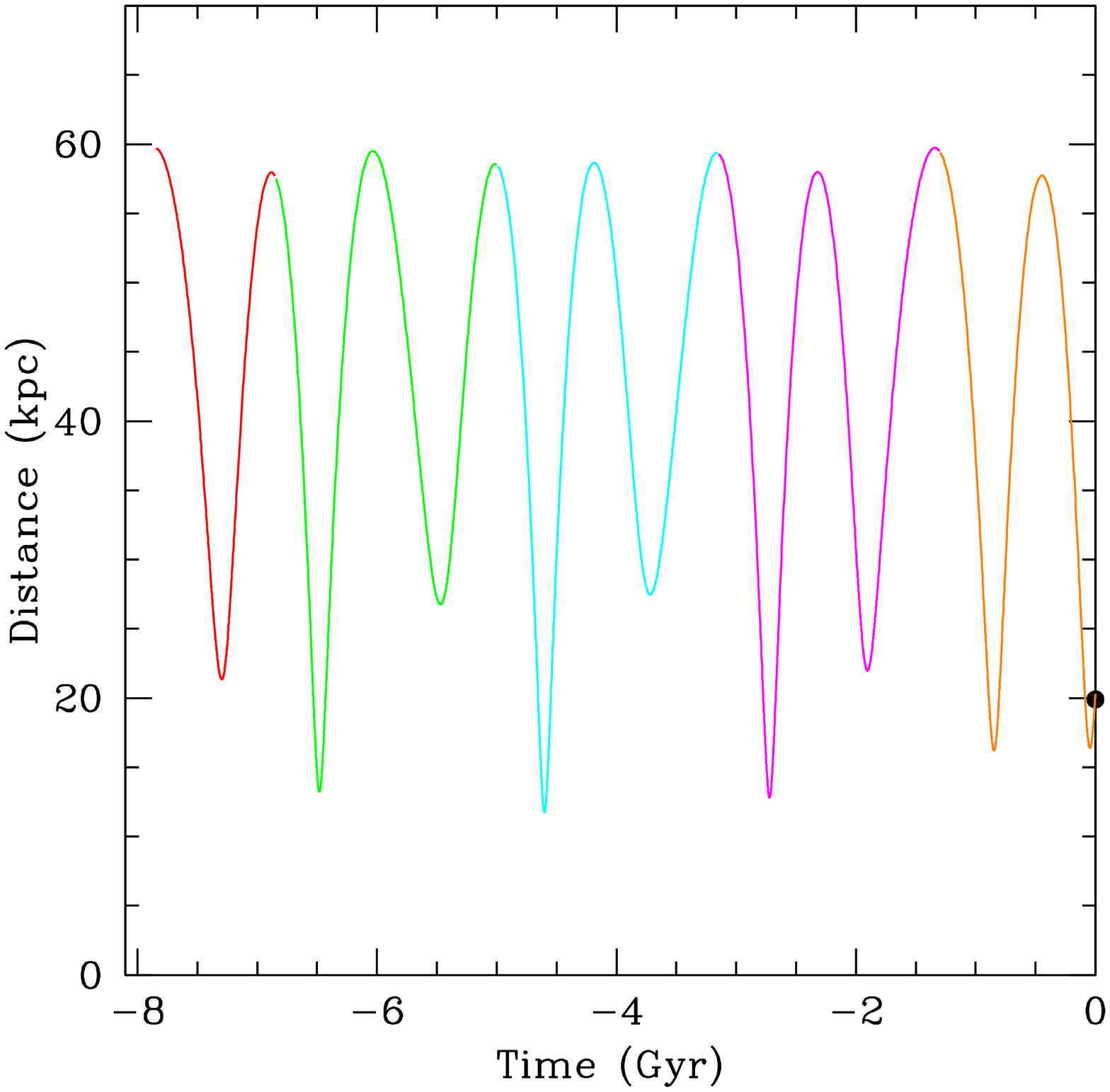}
\caption{Distance of the model Sgr dwarf from the Galactic center over its $\sim$ 8 Gyr orbital history.  
Time is in Gyr before the present day.  Different color-codings represent different orbital phases on which
debris was lost in the $N$-body model (i.e., orange debris on $N$-body plots throughout this paper correspond to debris lost on the orange colored orbits shown here, etc.).}
\label{OrbitPlot.fig}
\end{figure}

\begin{figure*}
\epsscale{0.8}
\plotone{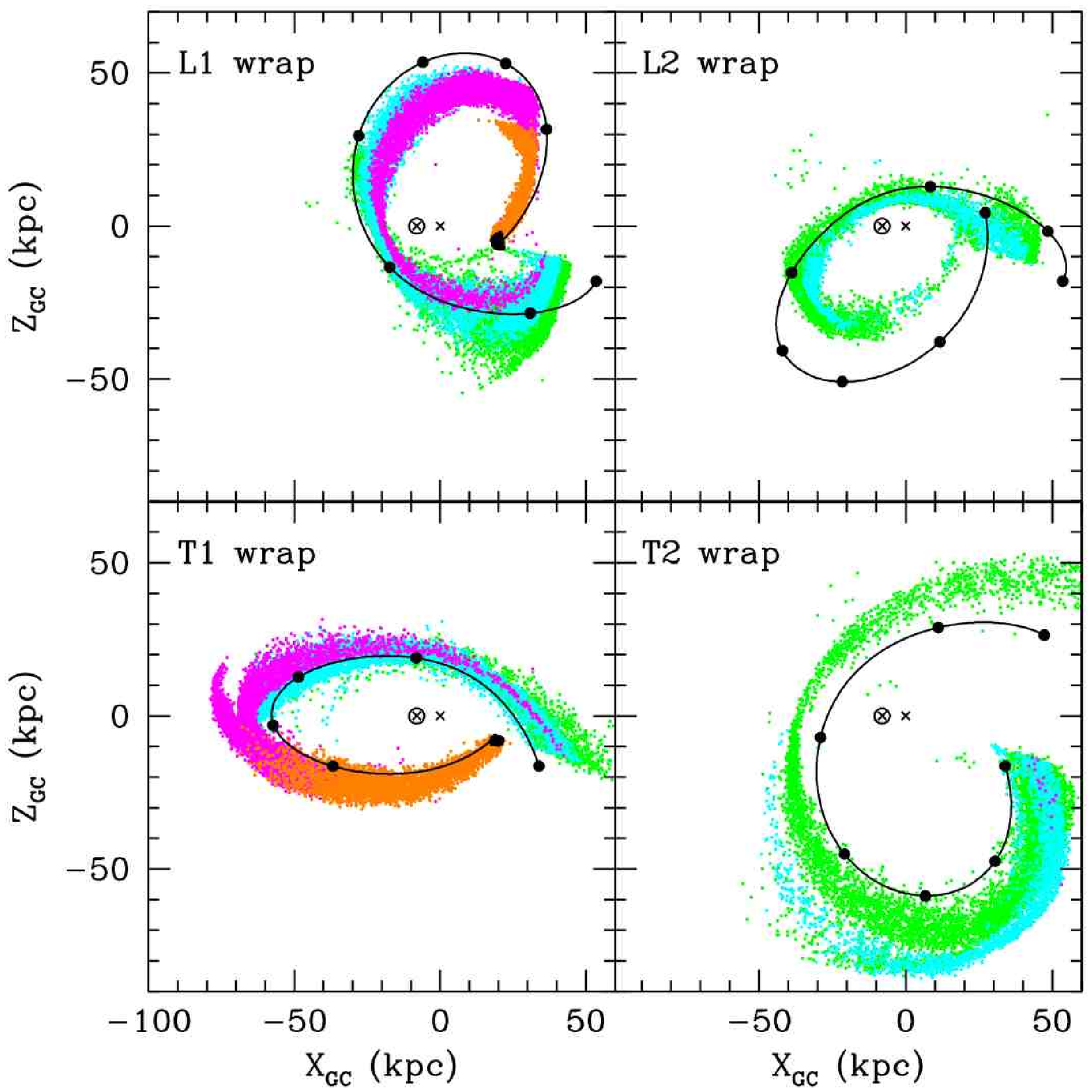}
\caption{\scriptsize Diagram showing the four wraps of the model Sgr stream in Galactocentric Cartesian coordinates $(X_{\rm GC}, Z_{\rm GC})$.
The L1/T1 wraps represent the primary leading/trailing tidal streams respectively, while the L2/T2 wraps represent the secondary leading/trailing tidal streams
of debris which has been wrapped $> 360^{\circ}$ around the Milky Way from Sgr.  Colors of individual debris particles represent the time at which
the particle became unbound from the Sgr dwarf (see Figure \ref{OrbitPlot.fig} for a key).  The solid black line represents the orbital path of the Sgr dwarf core, filled
black circles are drawn every 0.2 Gyr along the orbit.  The $\times$ indicates the location of the Galactic Center, while the circled $\times$ indicates the location of the Sun.}
\label{XYgrid.fig}
\end{figure*}

Since the Sgr streams frequently overlap themselves in all of the observable coordinates, 
it is useful to isolate specific wraps of the leading/trailing streams in order to discuss individual wraps more easily.
By sorting the streams according to angular momentum (relative to the Sgr core) and time at which a given debris particle became unbound from the model
satellite this is easily accomplished in the $N$-body simulations.
We color-code debris particles according to the time they became unbound from the satellite: 
Black points are currently bound to Sgr, orange points were stripped from the dwarf during the 
last two perigalactic passages (0 - 1.3 Gyr ago), magenta points during the previous two (1.3 - 3.2 Gyr ago), etc. (see Figure \ref{OrbitPlot.fig} for a detailed accounting
of the color-coding scheme).
As a convenient shorthand notation, we describe specific wraps of the 
streams as follows (see summary in Figure \ref{XYgrid.fig}):

\begin{description}

\item[L1:]  Primary wrap of the leading arm of Sgr (i.e., $0^{\circ} - 360^{\circ}$ angular separation from the dwarf).  Traced by orange/magenta points near the Sgr dwarf, magenta/cyan/green points at larger angular separations.

\item[L2:] Secondary wrap of the leading arm of Sgr (i.e., $360^{\circ} - 720^{\circ}$ angular separation from the dwarf).  Traced almost entirely by cyan/green points.


\item[T1:]  Primary wrap of the trailing arm of Sgr (i.e., $0^{\circ} - 360^{\circ}$ angular separation from the dwarf).  Traced by orange/magenta points near the Sgr dwarf, magenta/cyan/green points at larger angular separations.

\item[T2:] Secondary wrap of the trailing arm of Sgr (i.e., $360^{\circ} - 720^{\circ}$ angular separation from the dwarf).  Traced almost entirely by cyan/green points.

\end{description}

The major observational distinction of the present model from the oblate-halo model discussed by LJM05 is that the primary L1
wrap of the leading Sgr tidal stream is no longer predicted to pass closer than
$\sim 10$ kpc to the Sun; the center of the L1 streamer intersects the Galactic disk at a distance of 13.5 kpc in the direction of the Galactic anticenter
(see, e.g., Figures \ref{ModelSummary.fig} and \ref{XYgrid.fig}).
This is consistent with the absence of net vertical flow observed in the solar neighborhood in recent radial velocities surveys (Seabroke et al. 2008).

Due to the triaxial Galactic potential, the orbital period and pericenter/apocenter of Sgr varies from orbit to orbit (Figure \ref{OrbitPlot.fig});
the mean values of these parameters are 0.93 Gyr and 19/59 kpc respectively over the past 8 Gyr.
The present space velocity of Sgr in this model is ($U,V,W$) = (230, -35, 195) \kms.
For our assumptions about the distance to Sgr, the Local Standard of Rest,  and the solar peculiar motion this translates to
a proper motion of $(\mu_l \textrm{cos} b, \mu_b) = (-2.16, 1.73)$ mas yr$^{-1}$
($[\mu_{\alpha} \textrm{cos} \delta, \mu_{\delta}] = [-2.45, -1.30]$ 
mas yr$^{-1}$ in celestial coordinates) in the heliocentric rest frame.
This agrees within $1.9 \sigma$ with the proper motion measurements of Dinescu et al. (2005) who found 
$(\mu_l \textrm{cos} b, \mu_b) = (-2.35 \pm 0.20,  2.07 \pm 0.20)$ mas yr$^{-1}$  based on Southern Proper Motion Catalog values for 2MASS-selected giant star candidates,
and within $2.2 \sigma$ of the values $(\mu_l \textrm{cos} b, \mu_b) = (-2.62 \pm 0.22,  1.87 \pm 0.19)$ mas yr$^{-1}$ determined by Pryor et al. (2010).
Our tangential velocity for Sgr ($v_{\rm tan} = 251$ \kms) is also in good agreement with the measures of Irwin et al. (1996; $v_{\rm tan} = 250\pm90$ \kms)
and Ibata et al. (2001b; $v_{\rm tan} = 280 \pm 20$ \kms).
The mass of the Milky Way within 50 kpc in this model is $4.4 \times 10^{11} M_{\odot}$
(i.e., slightly smaller than that obtained by Sakamoto et al. 2003), corresponding to a halo scaling parameter
$v_{\rm halo} = 121.9$ \kms (see Equation \ref{haloeqn}).

The present mass of the Sgr dwarf in this model is $M_{\rm Sgr} = 2.5^{+1.3}_{-1.0} \times 10^8 M_{\odot}$; taking the
apparent magnitude of the bound core of Sgr to be $m_V = 3.63$ (Majewski et al. 2003) gives 
an absolute magnitude $M_V = -13.64$ and luminosity
$L_{\rm Sgr} = 2.4 \times 10^7 L_{\odot}$ for our adopted Sgr distance modulus ($[m-M]_0 = 17.27$; Siegel et al. 2007)
implying a mass/light ratio of $M_{\rm Sgr}/L_{\rm Sgr} = 10^{+6}_{-4} M_{\odot}/L_{\odot}$.  
This is at the low end of the range $M_{\rm Sgr}/L_{\rm Sgr} = 14 - 36 M_{\odot}/L_{\odot}$ 
estimated by LJM05; this difference reflects both our greater adopted distance to Sgr and our
choice to use the higher-resolution Monaco et al. (2007) measurement of the dispersion of the Sgr stream instead of the previous Majewski et al. (2004) value.
We note, however, that the derived mass is somewhat dependent on the structural model adopted for the Sgr dwarf.  
At present, the total interaction time of Sgr (and hence the initial mass of the satellite)
is relatively unconstrained: With reasonable assumptions for its initial mass the length of the tidal
streams observed in 2MASS and SDSS suggest that Sgr must have been disrupting for at least the last 3 Gyr (roughly corresponding to orange/magenta-colored
points in Figure \ref{ModelSummary.fig}).  We have opted to run our simulation for a significantly longer time ($\sim 8$ Gyr) in order to make concrete predictions about where Sgr
debris from earlier epochs may be expected; observational confirmation of the presence or absence of such older debris streams will provide a strong constraint on
the actual length of the interaction time.  Future models that more accurately treat the dark/light matter separation and are constrained to reproduce the bar-like core structure
of Sgr (e.g., {\L}okas et al. {\it in prep}.) may provide more stringent constraints on the $M/L$ ratio of the dwarf.

\subsection{The Nature of the Bifurcation}
\label{bifurc.sec}

\begin{figure*}
\epsscale{0.8}
\plotone{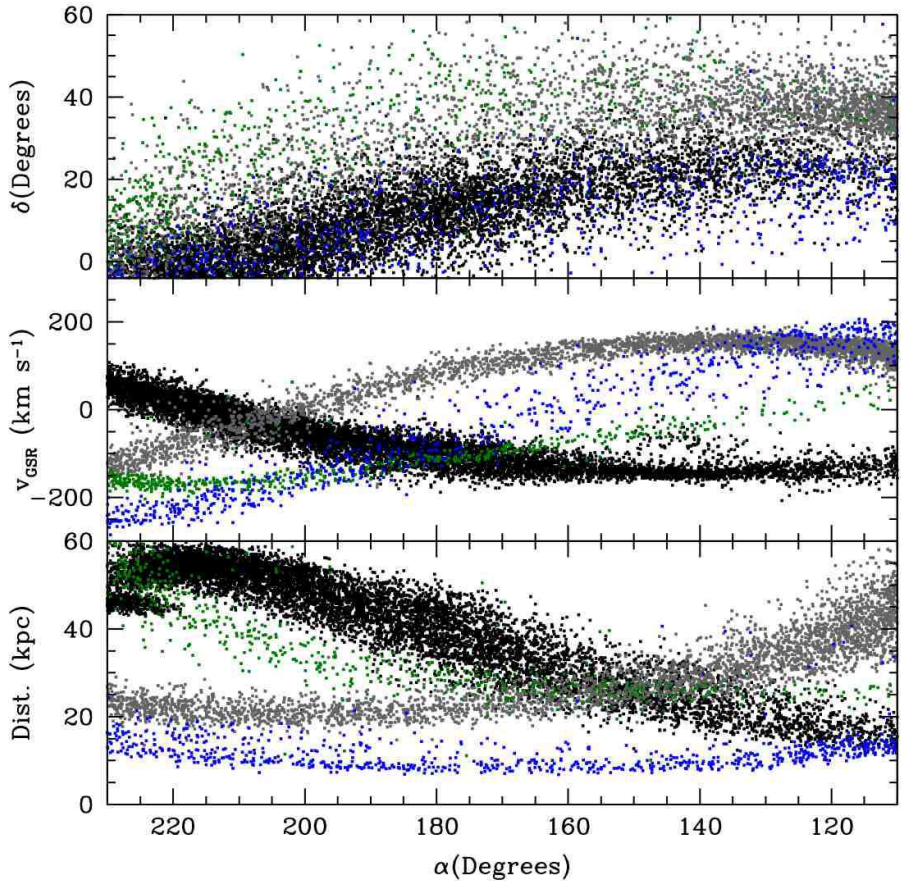}
\caption{\scriptsize Predicted locations, distances, and radial velocities of Sgr tidal debris in the `Field of Streams' region described by Belokurov et al. (2006).
This figure is shown in celestial $(\alpha,\delta)$ coordinates instead of the more natural $(\Lambda_{\odot}, B_{\odot})$ to aid
comparison to the figures of Belokurov et al. (2006) and Fellhauer et al. (2006).
Note that the color-coding in this figure is different than that adopted in all other $N$-body debris plots (e.g., Figure \ref{XYgrid.fig}):
Black/dark grey points represent L1/T1 leading/trailing arm debris, blue/dark green points 
represent L2/T2 leading/trailing arm debris.}
\label{bifurc.fig}
\end{figure*}

As indicated by Figure \ref{ModelSummary.fig}, our simulations match the high surface-brightness (hereafter HSB) track of the leading arm in the
 `Field of Streams' region described by Belokurov et al. (2006) but do not produce a bifurcation in the form of a low surface-brightness (hereafter LSB) branch
offset to higher declination.  Fellhauer et al. (2006) previously explained this bifurcation as the juxtaposition of the young ($< 4$ Gyr) L1 wrap of the leading arm
with the old ($> 7.4$ Gyr) T2 wrap of the trailing arm.\footnote{Note
that Fellhauer et al. (2006) call the equivalent of their L1/L2/T1/T2 streams the A/C/D/B streams respectively.}  This explanation is not viable for the current model:
As illustrated in Figure \ref{bifurc.fig}, 
the distance/radial velocity trends of the L1 (black points) and T2 (green points) streams are markedly different, in contrast
to Yanny et al. (2009) who demonstrate that the distance and radial velocity trends of the LSB and HSB streams
are indistinguishable.

However, we note that the Yanny et al. (2009) observations are not reproduced well by the Fellhauer et al. (2006) model either: This model predicts a factor $\sim 2$ difference
in distance between the LSB and HSB streams at $\alpha = 120^{\circ}$ that is not observed.
Additionally, given the strong evolution in the MDF with increasing separation $\Lambda_{\odot}$ from the Sgr dwarf (e.g., Chou et al. 2007) 
we should not expect the metallicity of the LSB and HSB streams to be similar (as observed by Yanny et al. 2009) if they represent extremely different dynamical
ages of tidal debris.  Certainly, if the LSB stream were significantly older than the HSB 
we should not expect the LSB stream to contain as many M-giants as the HSB stream since M-giants are a young tracer population.

$N$-body models of Sgr tidal destruction also predict that the young T1 trailing arm should pass through the Field of Streams region.  In both the model presented
here and the model described by Fellhauer et al. (2006), the T1 wrap traverses this region from $\alpha \sim 120^{\circ}$ to $\alpha \sim 220^{\circ}$
with a distance varying from $\sim 40$ kpc to $\sim 20$ kpc, and radial velocity varying from $\sim 125$ \kms to $\sim -125$ \kms.
Such tidal debris (grey points in Figure \ref{bifurc.fig}) is not observed in the Field of Streams, suggesting that the Sgr tidal streams are insufficiently long
to produce T1 trailing tidal debris in this region of sky (or at least, that the stellar density in the stream is declining sufficiently rapidly that it has not been detected in the SDSS).
This places a limit on the length of the Sgr tidal stream similar to that derived from other observational constraints that the Sgr tails represent $\lesssim 3$ Gyr
worth of tidal debris.  If the Sgr tails are so young, we should not expect any debris corresponding to earlier wraps (i.e., T2/L2) to exist and it is correspondingly impossible to interpret
the bifurcation as the juxtaposition of Sgr tidal debris torn from Sgr at significantly different times.

We therefore favor the explanation proffered by Yanny et al. (2009) that both the LSB and HSB streams may represent debris stripped from Sgr at similar times.
Indeed, we note that
models of Sgr orbiting in a non-spherical Galactic potential exhibit
precessional broadening of the angular width of the leading debris stream in this region
which manifests as  a
low surface brightness
`spray' to higher declinations similar to that of the LSB stream (this effect is increasingly noticeable for haloes with stronger departures from spherical symmetry).
This presents an alternative scenario wherein the bifurcation may arise due to substructure {\it within} the initial Sgr dwarf, such as a smaller dwarf satellite
that may have been bound to Sgr when it fell into the gravitational potential of the Milky Way
(see, e.g., Chou et al. 2009; Law et al. {\it in prep}), or simply to anisotropy of the stellar orbits within the
original dwarf which might be expected to give rise to tidal debris occupying only a subset of the angular width covered by our isotropic $N$-body model.

The detailed nature of the low surface brightness branch in the SDSS view of the Sgr stream is therefore unknown at present, 
and is not possible to reconcile with $N$-body models
that fit the unambiguous primary leading and trailing tidal streams illustrated in Figure \ref{ModelSummary.fig}.  While this feature may 
ultimately provide a constraint on the internal structure of the initial dSph satellite, exploration of such structure is beyond the scope of this contribution.


\section{Discussion: The Shape and Orientation of the Milky Way Dark Matter Halo}
\label{halodisc.sec}

Cold dark matter (CDM) models of galaxy formation generally predict that dark haloes should be mildly triaxial 
(e.g., Bullock 2002; Jing \& Suto 2002; Bailin \& Steinmetz 2005; Allgood et al. 2006; and references therein),
predominantly prolate  systems ($T_{\Phi} =  \frac{1 - (b^2/a^2)_{\Phi}}{1 - (c^2/a^2)_{\Phi}}  \gtrsim 0.66$; Allgood et al. 2006)
with characteristic central axis ratios\footnote{We adopt a convention in which axial ratios such as $c/a$ are given with a subscript of $\Phi$ ($\rho$) to indicate
that they are measured with respect to the potential (density) contours.}
  $(c/a)_{\Phi} \sim 0.72$, $(b/a)_{\Phi} \sim 0.78$ (Hayashi et al. 2007).  
While there is a tendency for haloes to be more spherical at larger radii (Allgood et al. 2006; Hayashi et al. 2007), the halo of Milky Way-like systems at a distance of $\sim 50$ kpc
is expected to be prolate with $T_{\Phi} \approx 0.88$ (based on the Via Lactea simulation of Diemand et al. 2007; Kuhlen et al. 2007).

Observationally, typical haloes appear to  be  rounder than predicted by collisionless CDM simulations,
as indicated by gravitational lensing mass models (e.g., Hoekstra et al. 2004; Mandelbaum et al. 2006; Evans \& Bridle 2009), 
polar ring galaxies (e.g., Iodice et al. 2003), and X-ray isophotal studies  (e.g., Buote et al. 2002; although c.f. Diehl \& Statler 2007; Flores et al. 2007).
This possible mismatch is perhaps unsurprising because the growth of baryonic structures within these haloes tends to make them 
considerably rounder (e.g., Debattista et al. 2008; Valluri et al. 2009), inflating
the principal axis ratios by as much as $\sim 0.2 - 0.4$ within the virial radius (Kazantzidis et al. 2004).
However, it is difficult to constrain the actual triaxiality of individual haloes because galaxies 
are projected onto the two-dimensional plane of the sky 
and observations are relatively insensitive to the mass distribution along the line of sight (although c.f. Corless et al. 2009).

It is therefore noteworthy that for the only galaxy  where we have a fully three-dimensional  view from the interior of the mass distribution (i.e., the Milky Way)
it appears (\S \ref{combinedconst.sec}) that the best-fitting axial ratio at radii $20 < r < 60$ kpc describes an almost perfectly {\it oblate} ellipsoid
with minor/major  axis ratio $(c/a)_{\Phi} = 0.72$, 
intermediate/major axis ratio $(b/a)_{\Phi} = 0.99$, triaxiality $T_{\Phi}  = 0.06$, and the minor axis located in the plane of the Galactic disk
in the direction $(l, b) = (7^{\circ}, 0^{\circ})$.
In this section we discuss the physical interpretation of these results, along with possible effects that could influence our derivation of the halo shape.

\subsection{Physical Validity of the Triaxial Halo Model}


\subsubsection{Tracing the Underlying Mass Distribution}
\label{massmodel.sec}

In \S \ref{model.sec} we introduced flattening directly into the contours of the gravitational potential because this was analytically tractable
and consistent with the method of previous studies (in particular LJM05).  
It is  well-known however that as the ellipticity of the potential becomes more extreme, the corresponding mass distribution required to generate it
may become unphysical (i.e., have regions of negative mass).
In order for any model of the Galactic halo to be believed, it must correspond to a physically realizable density distribution.

\begin{figure*}
\plotone{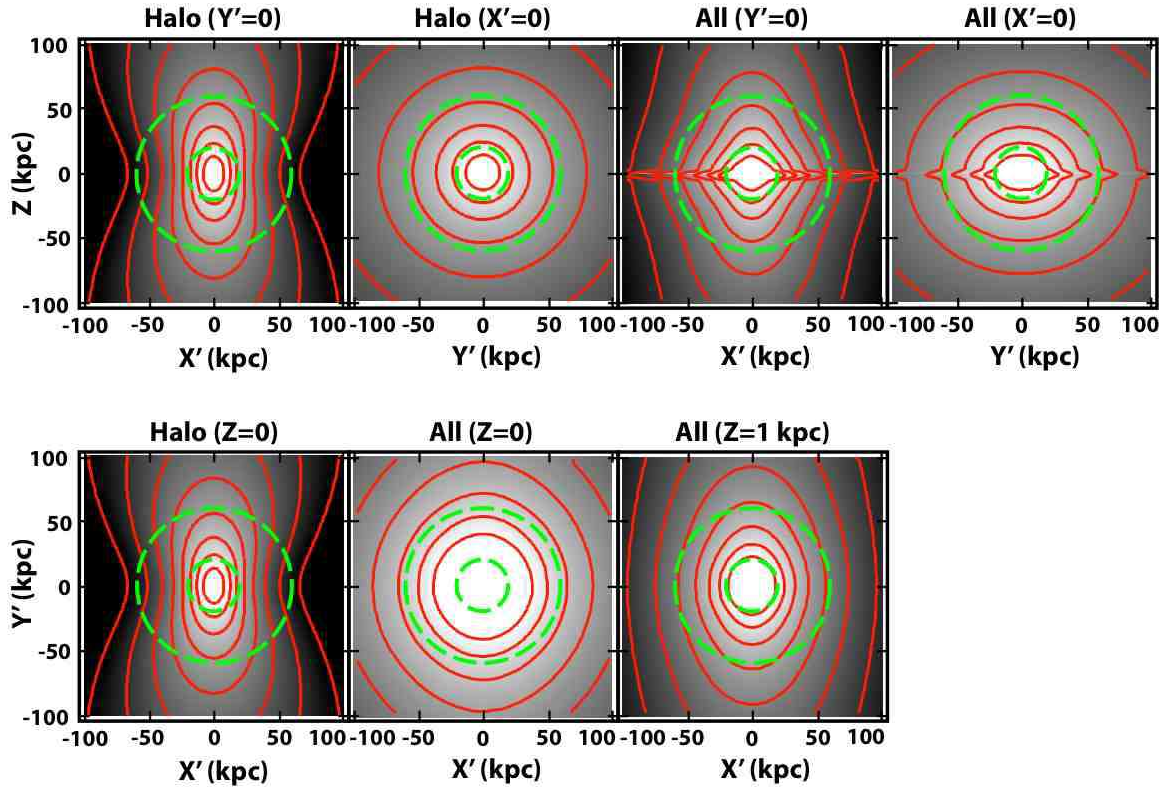}
\caption{Density maps viewed along different axes for the best-fit Galactic halo model in an $(X', Y', Z)$ coordinate system aligned with the principal axes of the halo.  
Density maps are shown for the dark halo alone, and for the sum of all mass components (i.e., logarithmic halo, Miyamoto-Nagai disk, and Hernquist spheroid).
The greyscale/red contours are logarithmic and with the same scale in all panels.  Note that while
the axisymmetric density profile of the disk dominates in the disk plane ($Z=0$), at only 1 kpc above the plane
the halo contribution produces noticeably elliptical contours in the $X'-Y'$ plane at radii $r \gtrsim 20$ kpc.
The dashed green circles indicate the mean apocenter/pericenter of the Sgr orbit.}
\label{density.fig}
\end{figure*}

In Figure \ref{density.fig} we show the corresponding contours of the mass distribution underlying our best-fit model for the Milky Way (as calculated by taking the Laplacian of the
gravitational potential).  As expected, the mass distribution is less spherical than the isopotential surface;
the contours of the dark matter halo are approximately elliptical at small radii (i.e., within $\sim 40$ kpc) with $(c/a)_{\rho} = 0.44$, $(b/a)_{\rho} = 0.97$, and $T_{\rho} = 0.07$.
Although some pinching of the dark matter density contours is apparent at larger radii the mass distribution is everywhere positive.
Figure \ref{density.fig} also demonstrates that while addition of the baryonic mass component significantly changes the net density distribution at small radii and near the Galactic disk plane,
the density distribution at larger radii probed by the orbit of Sgr is dominated by the dark halo component.

We additionally confirmed the validity of our results by constructing 
a triaxial NFW (Navarro et al. 1996) halo model in which the axial flattenings were introduced in a more physically motivated manner directly into the
mass distribution and where the  resulting accelerations were computed using the approximate form of the corresponding potential given by Lee \& Suto (2003).
As expected, there are slight differences in the path of satellites orbiting in this NFW halo and in our best-fit logarithmic potential, but these 
differences are minor and likely due to the fact that we did not fine tune the NFW model.

\subsubsection{Orientation of Principal Axes}

While the mass distribution underlying our best-fit model of the Galactic halo is physically reasonable, the {\it orientation} of the halo (i.e., strongly non-axisymmetric in
the plane of the Galactic disk, with the angular momentum vector of the disk aligned with the intermediate axis of the halo) is unexpected because
orbits about intermediate axes tend to be unstable (e.g., Adams et al. 2007).
This is not a concern for stellar orbits in the Galactic disk
because the gravitational potential at small radii $r \lesssim 20$ kpc is dominated by the (nearly) axisymmetric baryonic mass distribution
rather than the triaxial dark halo.
Indeed, preliminary results of orbit integration in the gravitational potential derived here indicate that loop orbits are permitted, suggesting that
a self-consistent disk may still be supported (R. Johnson and K.V. Johnston, private communication, 2010).
However, this situation begs the question of why the Galactic disk formed in its present orientation
unless the principal axes of the dark halo either 
twist substantially with radius (in contrast to the general prediction of Hayashi et al. 2007)
or evolve with time.
While we cannot rule out the possibility that rotation of the dark halo has contributed in some regard to our results, we neglect such effects because typical haloes
are expected to rotate extremely slowly (e.g., Bullock et al. 2001a; Herbert-Fort et al. 2007, 2008; Bett et al. 2009).

Typically, simulations of disk galaxy formation suggest (e.g., Debattista et al. 2008) that the minor axes of the dark halo
and the baryonic disk should be approximately aligned (to within an average of $\sim 30^{\circ}$; see discussion by Warren et al. 1992; van den Bosch 2002;
Bett et al. 2009) because both are formed from  accreted matter with similar specific angular momenta.
If a disk were to form in an elliptical halo potential such as that proposed in \S \ref{combinedconst.sec}, simulations suggest that the disk too should become
elongated and non-axisymmetric
(see, e.g., Gerhard \& Vietri 1986; Debattista et al. 2008).
It will therefore be intriguing to investigate whether the relative orientation of the host dark matter halo and baryonic disk proposed here
can be strongly motivated within the current CDM paradigm.

It is worth asking as well whether such an inclined oblate halo can explain the observed distribution of Galactic satellites, which simulations suggest (Zentner et al. 2005)
should be distributed along the major axis of the host halo (i.e., 
have a pole aligned with the minor axis).
The Fornax-Leo-Sculptor great circle has a pole $(l,b) = (135^{\circ}, -2.9^{\circ})$ (Lynden-Bell 1982; see also Kunkel \& Demers 1976; Majewski 1994; Fusi Pecci et al. 1995), 
and a recent update (Metz et al. 2007, 2009)
including the faint new dwarfs discovered in the SDSS gives a revised satellite pole $(l,b)=(159.7, -6.8)$.  The updated pole of Metz et al. (2009) lies within $\sim 30^{\circ}$
of the minor axis we derive for the Galactic halo, but it is difficult to determine whether this similarity is simply an unrelated coincidence.\footnote{If the probability distribution
of the satellite pole were uniform across the entire sky, the chance for it to be coincidentally aligned with $(l,b) = (7^{\circ}, 0^{\circ})$
to within $30^{\circ}$ is $\sim 13$\%.}



\subsection{Alternatives to the Triaxial Halo Model}

In LJM05, we conducted an extensive parameter space search in an effort to find a combination of 
Milky Way + Sgr parameters capable of simultaneously matching both the observed angular position and distance/radial velocity
trends of leading tidal debris.  In particular, we explored 
the effect of varying
the proper motion of Sgr ($v_{\rm tan}$), the distance to Sgr ($D_{\rm Sgr}$),  the distance from the Sun to the Galactic Center ($R_{\odot}$),
the total mass of the Galactic disk (via the scaling parameter $\alpha$ in Equation \ref{diskeqn}),
 the dark halo scale length ($d$), the dark halo flattening perpendicular to the disk ($q_z$),
and the total mass scale of the Milky Way as parameterized by the rotation speed of the Local Standard of Rest ($v_{\rm LSR}$).

\begin{figure*}
\plottwo{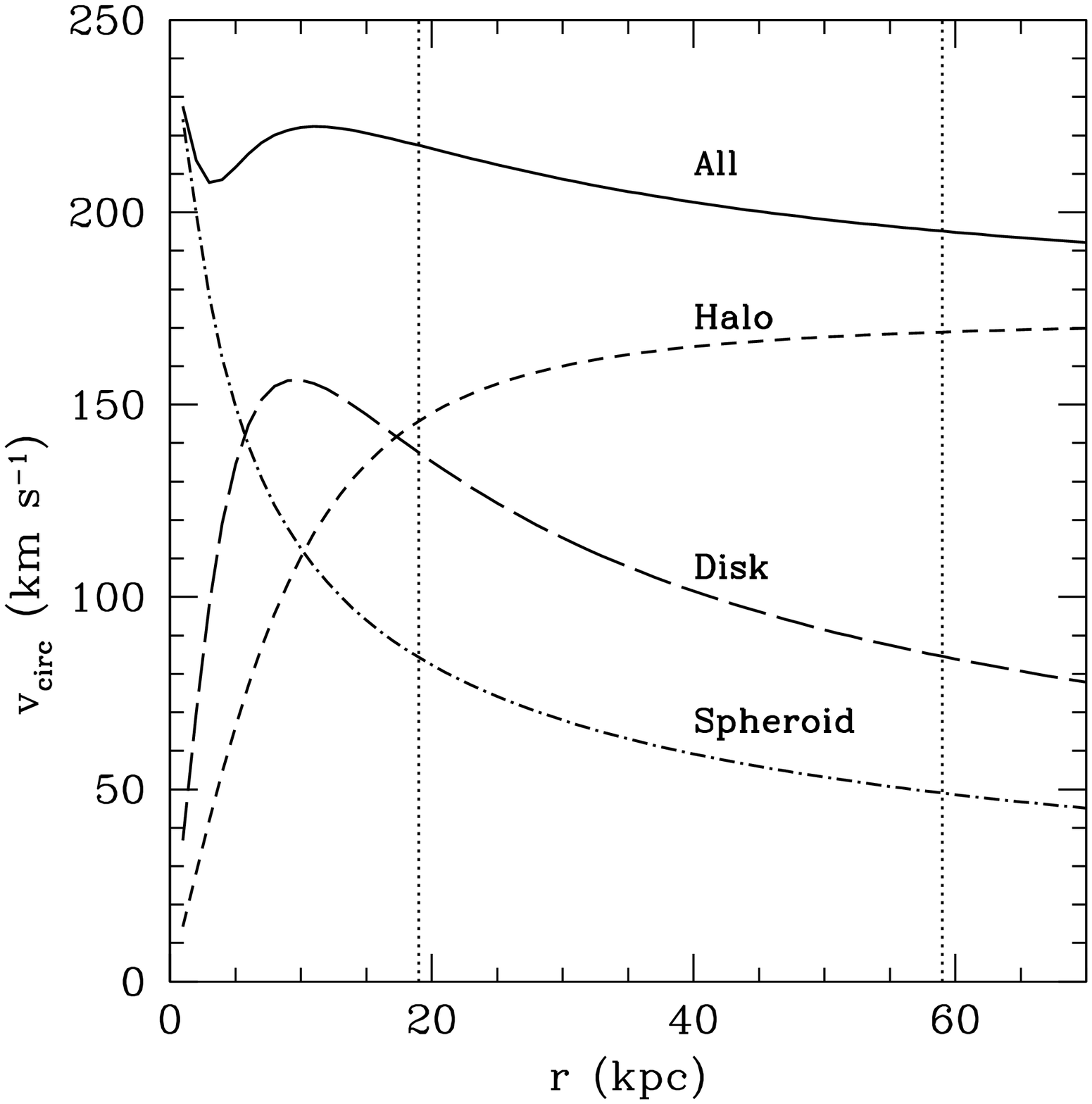}{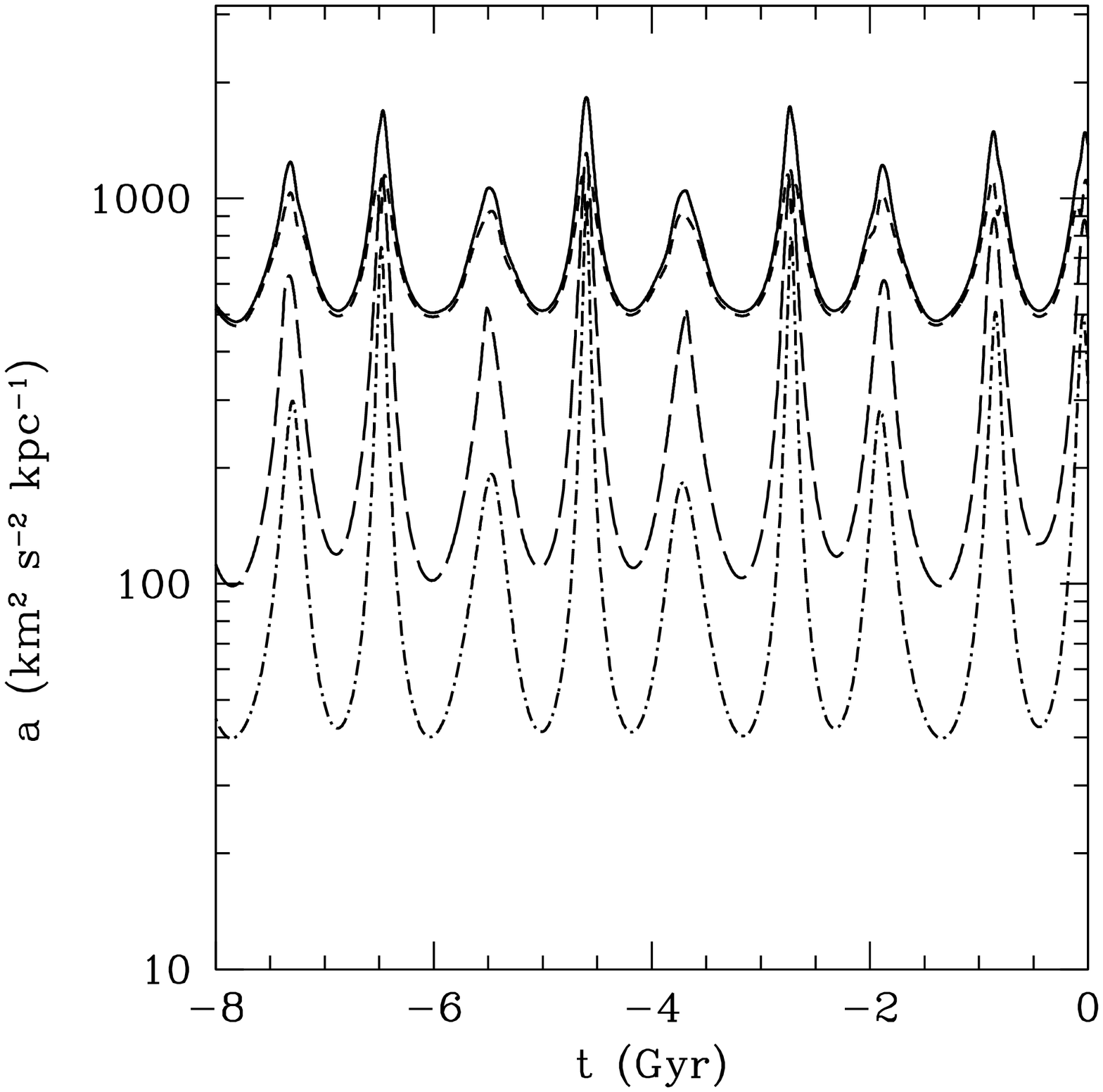}
\caption{Left-hand panel: Contribution of the dark halo, stellar disk, and stellar spheroid to our model Galactic rotation curve as a function of radius
in the Galactic disk plane
(for purposes of illustration the halo is taken to be axisymmetric with $q_1 = q_2 = 1.0$).  The vertical dotted lines
denote the mean pericenter/apocenter of the Sgr orbit.  Right-hand panel: Acceleration of the Sgr dwarf in the gravitational potential of the Milky Way
as a function of time over its assumed 8 Gyr orbital history.
The dark halo, stellar disk, stellar spheroid, and total contributions to the rotation curve and acceleration of Sgr are indicated by dashed, long-dashed, dot-dashed, and solid
lines respectively in both panels.}
\label{rotcurve.fig}
\end{figure*}

Although the Galactic disk contributes significantly to the overall flattening of our adopted Galactic potential at small radii (e.g., Figure \ref{density.fig}), it has 
only a minor effect at the range of radii ($r \sim 20 - 60$ kpc) probed by the orbit of Sgr.  As demonstrated by Figure \ref{rotcurve.fig}, the dark halo
in our model is
the dominant contributor to the total gravitational potential at these radii. 
Indeed, if we decrease the mass of the Galactic disk by 50\% (by setting $\alpha = 0.5$
in Equation \ref{diskeqn}) and increase the total halo mass correspondingly (so that the Local Standard of Rest remains constant at
$v_{\rm LSR} = 220$ \kms), we can repeat the parameter space search described in \S \ref{mwprop.sec}.  Scaling the proper motion $v_{\rm tan}$ of Sgr to again
best-fit the trailing-arm velocity trend, we find optimal values for $q_1, q_z, \phi$ almost identical to those derived previously (although the overall $\chi^2$ of
the best-fit model is not as good as for the case $\alpha = 1.0$).
We are therefore confident that the details of the Galactic disk model do not significantly bias our results.


Similarly, the details of the triaxial Galactic bar (which we did not include in our model) and stellar halo are unlikely to change our results.
Neither are sufficiently massive to have a great effect on the gravitational potential at distances $r \gtrsim 20$ kpc, 
and neither have  orientations akin to that which we derive for the dark halo.
The major axis of the bar is thought to lie within $\sim 15^{\circ}-20^{\circ}$ of the $X$-axis (e.g., Blitz \& Spergel 1991;
Nakada et al. 1991; Morris \& Serabyn 1996; Babusiaux \& Gilmore 2005), similar
to the minor axis of the dark halo.  While the stellar halo is significantly 
triaxial ($[c/a]_{\rho} \approx 0.65$, $[b/a]_{\rho} = 0.75$, $T_{\rho} = 0.76$; Newberg \& Yanny 2006; see also Larsen \& Humphreys 1996; Xu et al. 2006)
with a major axis lying approximately in the Galactic plane ($\sim 50-70^{\circ}$ from the $X$ axis) the minor axis is not apparently contained within the $X-Y$ plane but 
located $\sim 13^{\circ}$ from the $Z$ axis.

Throughout the entirety of the LJM05 parameter space search, the only parameter that was found to have a 
significant effect
on the angular position/distance/radial velocity trend of the Sgr leading arm was the flattening of the dark halo $q_z$.
Although  no single choice of $q_z$ can simultaneously
reproduce all observational constraints, we have demonstrated (see also LJM09) that a fully triaxial model in which the short axis is approximately aligned
with the Galactic $X$ axis is capable of doing so. 

Such a triaxial halo is therefore obviously attractive in the sense that {\it only} $N$-body models in such a halo succeed in modeling the Sgr -- Milky Way system where previous
efforts (e.g., Helmi et al. 2004; LJM05; Fellhauer et al. 2006; Mart{\'{\i}}nez-Delgado et al. 2007) have failed.
Despite the extensive exploration of parameter space undertaken by ourselves (this paper; LJM05; LMJ09) and other groups
(e.g., Helmi et al. 2004; Fellhauer et al. 2006; Mart{\'{\i}}nez-Delgado et al. 2007) however, the space explored to date is far from exhaustive
and it is possible that the triaxial halo model may prove to be simply a numerical
crutch that mimics the effect of some as-yet unidentified alternative.  
In the following sections, we briefly discuss a few possibilities that may obviate the need for triaxiality to explain the extant data.

\subsubsection{Alternative Formulations for the Gravitational Potential}

While we have focused on a logarithmic formulation of the Galactic gravitational potential, other forms could also reasonably be adopted.
In LJM05 we explored axisymmetric NFW (Navarro et al. 1996) models and were unable to resolve the halo conundrum.  As demonstrated in \S \ref{massmodel.sec}
however, fully triaxial NFW models can give rise to orbits for Sgr that match  the observed angular precession and distance/radial
velocity trends of leading tidal debris.  We find that orbits within logarithmic and NFW haloes are sufficiently similar that there is no reason to differentiate between
the two for our present purposes.

Fellhauer et al. (2006) claim that a mass distribution of the form specified by Dehnen \& Binney (1998) can produce some effects similar to prolateness in a logarithmic halo.
However, using such a model these authors were nevertheless unable to reproduce the distance trend of leading tidal debris as seen in the SDSS (Belokurov et al. 2006)
and concluded that their Set D of proper motions for Sgr (i.e., those derived by LJM05) in a logarithmic halo provided the best overall fit to the observational
data then available.  We therefore find no reason to prefer a Dehnen \& Binney (1998) model over the triaxial logarithmic halo model presented here.

\subsubsection{Evolution in the Orbit}

Although our model dSph has been orbiting in the Galactic potential for $\sim$ 8 Gyr, we have not considered the influence of any evolution in either the orbit of Sgr or the underlying
gravitational potential of the Milky Way.
We expect there was actually evolution in both the depth, and the {\it shape} of the potential (e.g., Kuhlen et al. 2007).
While we expect the streams to be independent of past evolution in the potential of the Milky Way (since tidal streams respond adiabatically to changes in their host
potential; see discussion by Pe{\~n}arrubia et al. 2006) 
some evolution has certainly occurred in the orbit of Sgr since
it fell into the Milky Way.
Dynamical friction and other such evolutionary effects therefore reflect an uncertainty in our models, although we note that (as discussed by LJM05)
such effects alone are unlikely to be able to resolve the halo conundrum.  Further investigation of these effects 
must (among other things) properly account for the relative distribution and mass-loss history of both the dark and baryonic matter in Sgr,
and is therefore beyond the scope of this contribution.

\subsubsection{Dwarf --- Tail Gravitational Interaction}

As discussed by Choi et al. (2007), the gravitational influence of the bound satellite on debris in the tidal tails can shift the trajectories
of actual tidal debris away from the orbital path traced by the bound core of the dSph.  For sufficiently massive satellites (greater than $\sim 0.1$\% the virial mass of the host), 
this effect can be appreciable and  degenerate with that caused by flattening
(or triaxiality) in the underlying gravitational potential of the Milky Way.
However, we note that despite claims to the contrary 
(Choi et al. 2007) our simulations fully account for this effect; the self-gravity of the bound satellite is applied to all $N$-body particles at all
times regardless of whether or not they are in the Sgr core or in the tidal tails.

\subsubsection{Gravitational Influence of the Magellanic Clouds}


The Large Magellanic Cloud (LMC) is $\sim 50$ kpc distant in the direction $(l,b) = (280.5^{\circ}, -32.8^{\circ}$) 
(van der Marel et al. 2002), placing it within $\sim 0.5$ kpc of the Galactic $Y-Z$ plane.
Based on great-circle fits to the Magellanic stream, the pole of the system appears to be $(l_{\rm p},b_{\rm p}) = (188.5^{\circ}, -7.5^{\circ}$) (Nidever et al. 2008).
That is, the pole of the Magellanic stream is aligned with the short axis of the Galactic dark halo as derived in this contribution to within $\sim 1^{\circ}$ in Galactic 
longitude.\footnote{They differ by
$7.5^{\circ}$ in Galactic latitude, but the symmetry axes of the dark halo were fixed to lie at $b =0^{\circ}$.}
This suggests two possibilities: 
(1) The Magellanic Clouds may have fallen into the Milky Way along the plane of the long and intermediate axes of the dark matter halo (i.e., the alignment is a {\it consequence} of the
dark matter distribution), or 
(2) the Magellanic Clouds may exert a significant gravitational acceleration on the Sgr dwarf, with the mass of the Clouds at large radii in the 
$Y-Z$ plane mimicking the effect of an extended dark matter distribution in this plane (i.e., the alignment is the {\it cause} of the apparent dark matter distribution).

\begin{figure*}
\epsscale{0.8}
\plotone{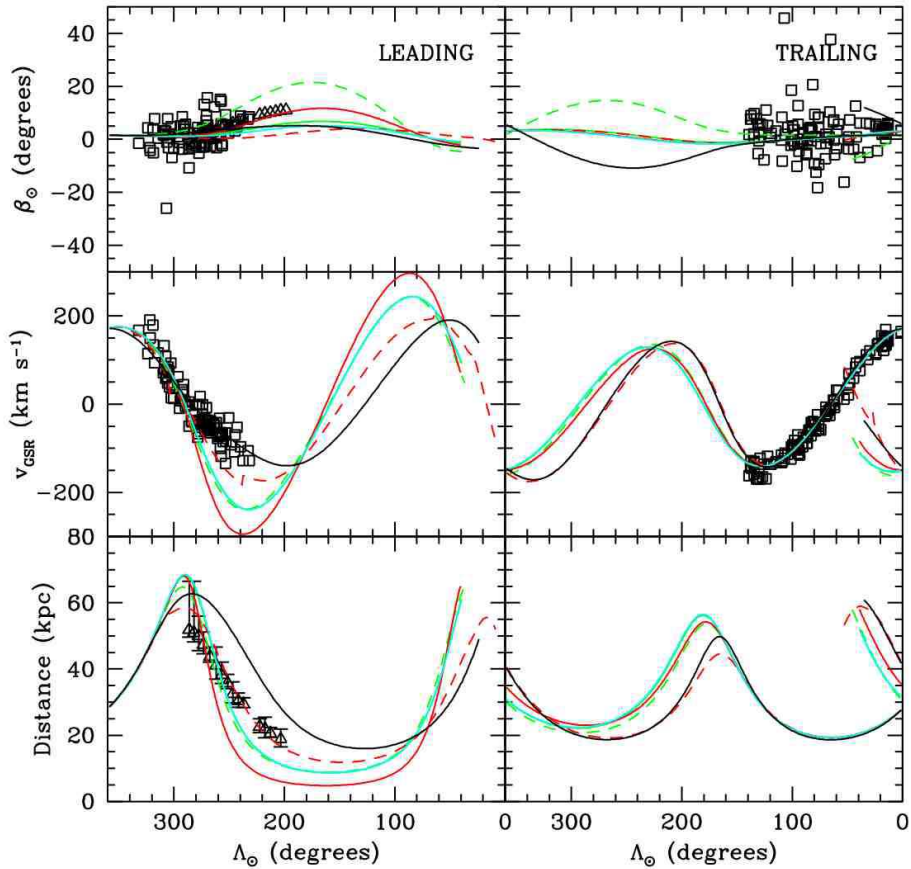}
\caption{\scriptsize Effect of the LMC on the angular path, distance, and line of sight velocity of the leading/trailing Sgr orbit (left- and right-hand panels respectively).  
Observational data (open symbols) are as in Figure \ref{ModelSummary.fig}.  
The cyan line represents the path of the Sgr dwarf in a simple spherical halo model, green solid/dashed lines the path in a spherical halo with a point-mass LMC that is
1\%/10\% of the Milky Way mass, and red solid/dashed lines the path in a spherical halo with a ring-model LMC that is 1\%/10\% of the Milky Way mass.  It is difficult
to see the solid green line in places since the cyan line lies almost on top of it.
Note the strong deviations of the models incorporating the LMC from the predictions of a simple spherical halo model, especially for larger values of the LMC mass.
The black line indicates the path of the dwarf in the triaxial halo model for comparison.}
\label{LMCeffect.fig}
\end{figure*}

A precise description of the second scenario 
would require a comprehensive model for the orbital and mass-loss history of the Magellanic Clouds, but it is possible to understand the magnitude of their
gravitational influence on the path of Sgr tidal debris by
integrating massless test-particles along the Sgr orbit in a Milky Way with a spherical dark halo and simplified model of the LMC.
We first test the effect of introducing a fixed point-mass into the gravitational potential of the Milky Way at the present location of the LMC, 
whose mass is allowed to vary from $1$ - 10\%
of the total mass of the Milky Way within 50 kpc (i.e., $M_{\rm LMC} = 6 \times 10^9 M_{\odot} - 6\times 10^{10} M_{\odot}$).
As illustrated in Figure \ref{LMCeffect.fig}, a  fixed LMC with a mass 10\% that of the Milky Way can produce a significant deviation on the order of $20^{\circ}$ 
in the orbital path of leading Sgr tidal debris.

Of course the LMC is not fixed in space, but describes an orbit about the Milky Way confined approximately to  the $Y-Z$ 
plane.  We therefore develop a second, slightly more realistic model in which the LMC
is described in a time-integrated sense by a ring of mass at a radius of 50 kpc in the $Y-Z$ plane.  The gravitational acceleration at an arbitrary point in space resulting from such a configuration is
described using an analytical approximation to the numerically evaluated elliptical integrals (see derivation for the electrostatic case by Zypman 2005).
This second model produces even more noticeable deviations from the original orbital path, with appreciable shifts in both the leading arm angular position and radial velocities
expected for models in the range of masses $1 - 10$\% that of the Milky Way.

The detailed effect of the Magellanic Clouds on the orbit of Sgr is extremely uncertain.  Both our point-mass and ring-mass models are obviously oversimplifications, neglect
the SMC entirely, and assume that the gravitational
influence of the LMC has been fixed over the  entire interaction history of the 
Sgr dwarf ($\sim 8$ Gyr in the current model).  Since the Clouds are currently thought to be near the pericenter of their orbit, for the majority of the lifetime of Sgr they likely lay at greater distances
where their gravitational force would be smaller.  Indeed, some recent lines of investigation suggest that they may have only recently fallen into 
the gravitational potential of the Milky Way for the first time (e.g., Besla et al. 2007), drastically limiting the time during which they could have torqued the orbit of the Sgr dwarf.
Of the various orbital paths shown in Figure \ref{LMCeffect.fig}, only the triaxial halo model actually succeeds in reproducing all of the observational data.  However, given
the myriad uncertainties in the Milky Way - LMC - Sgr system the important conclusion to draw
is that gravitational peturbations from the Magellanic Clouds {\it  may} be sufficient to torque the leading arm of the Sgr stream significantly,
especially if the effect is enhanced by the response of the  Milky Way dark matter halo itself to the passage of the Clouds (see, e.g., Weinberg 1998).

\subsubsection{Alternative gravity}

Modified Newtonian gravity (MOND; see Milgrom 1983; Bekenstein 2004) 
on Galactic scales has  been proposed as an alternative theory to explain the flat rotation curves typical of galaxies without recourse to an extended halo of dark matter.
Read \& Moore (2005) demonstrated that MOND was an equally viable alternative to the oblate dark matter halo 
model presented by LJM05, capable of fitting
the angular position of the leading tidal debris but not the radial velocities.
Now that a dark matter model has been able to successfully match {\it both} observational constraints, 
it is worthwhile to investigate whether a similarly successful MOND model can be constructed.
Since the key to resolving the halo conundrum appears to have been the inclusion of a strongly non-axisymmetric component to the gravitational potential,
it is not immediately obvious how such a potential could be produced in a MOND model by 
the predominantly axisymmetric distribution of baryonic matter typically adopted for the  Milky Way
(although c.f. Wu et al. 2008; Widrow 2008).



\subsection{Testing the Model}

Fortunately, the model presented here for the shape, size, and orientation of the Galactic gravitational potential is easily testable; it must be 
capable of explaining the orbital paths of other Galactic satellites that similarly  provide constraints on the mass distribution in the Milky Way.
Typically, previous studies of tidal debris systems in the Galactic halo
have only explored the effects of changing the axial scalelength perpendicular to the Galactic disk (i.e., $q_z$), and it is unclear whether
satisfactory solutions for other satellites can be found in such a strongly non-axisymmetric potential as that suggested here.
In particular, can realistic orbits be found to match such systems as the Monoceros tidal stream (e.g., Pe{\~n}arrubia et al. 2005),
the Magellanic Clouds\footnote{While simulations generally favor haloes which are extended  along the Galactic $Z$ axis (similar to those in the triaxial model),
preliminary results indicate that the Magellanic Clouds are relatively insensitive to the distribution of mass along the Galactic $Y$ axis and can neither confirm
nor deny the triaxial model presented here (G. Besla, priv. comm.).}
(e.g., Besla et al. 2007; R{\u u}{\v z}i{\v c}ka et al. 2007), the ``orphan stream'' (Belokurov et al. 2007), the GD-1 stream (Koposov et al. 2009),
the newfound Cetus Polar Stream (Newberg et al. 2009),
and tidally disrupting Galactic globular clusters such as Pal 5 (Odenkirchen et al. 2009)?
Similarly, is a triaxial Galactic halo consistent with observations of the Galactic HI disk (e.g., Olling \& Merrifield 2000; Kalberla et al. 2005; and references therein)?
It is intriguing that  Saha et al. (2009) discuss evidence to suggest that the dark matter halo must be similarly non-axisymmetric in the Galactic $XY$ plane
in order to explain the asymmetric flaring of the disk (although c.f. Narayan et al. 2005).
Future observations of hypervelocity stars at large distances may also provide key diagnostics (e.g., Gnedin et al. 2005; Yu \& Madau 2007; Perets et al. 2009).


\section{Explaining the Evolution of the MDF along the Tidal Streams}
\label{feh.sec}

Recently, numerous measurements have been made (e.g., Vivas et al. 2005; Chou et al. 2007, 2009; Monaco et al. 2007; Yanny et al. 2009; Starkenburg et al. 2009) 
of the metallicity distribution function (MDF) along the Sgr stream, although
a uniform picture of the evolution of the MDF  is difficult to construct because
systematic differences in the metallicity sensitivity between stellar populations
traced by various surveys can bias
the resulting trends.    However, significant evidence is starting  to emerge for a gradient of the MDF with Sgr orbital longitude.
Bellazini et al. (2006) for instance observed that the ratio of blue horizontal branch to red clump stars changes between the main body of Sgr and the stellar streams
(demonstrating an age/metallicity gradient), and similar claims of such evolution have been made by Monaco et al. (2007)
who found that while the core of Sgr has a mean metallicity $\langle \feh \rangle = -0.35 \pm 0.19$ the tidal streams
have a mean metallicity of  $\langle \feh \rangle = -0.70 \pm 0.16$.
The most comprehensive picture presented to date however
has been that of Chou et al. (2007), whose uniform selection technique demonstrates that the MDF of the Sgr M-giant population continuously evolves from 
a median $\feh \sim -0.4$ in the Sgr core to $\feh \sim -1.1$ dex over the $\sim 180^{\circ}$ length of the leading tidal arm.

In this section, we attempt to reproduce the gradient in the MDF
by making physically motivated assumptions about the star formation history of the Sgr dwarf.
We consider only the M-giant population observed by Chou et al. (2007) and Monaco et al. (2007) in order to ensure a uniform selection function at all orbital longitudes.
While Chou et al. (2009) have recently extended their earlier work (Chou et al. 2007) by additionally measuring the evolution in the abundance
patterns of  titanium (Ti), yttrium (Y), and lanthanum (La), we do not incorporate these additional elements into our analysis.

\subsection{Enrichment Model}
\label{enrichmodel.sec}

Generally, we expect that star formation in Sgr is likely to have occurred near the bottom of its gravitational potential well, and that stars thus formed in the
core would gradually diffuse outwards with time.  Remnants of such stellar populations would tend to persist in the Sgr core until the present day 
and we therefore assume
that the stellar populations present in the Sgr streams correspond to counterparts
in the Sgr core.
These Sgr core populations have been traced using high-precision {\it HST}-ACS photometry by Siegel et al. (2007), who
find five major star formation episodes:

\begin{description}

\item[Population A:] M54 metal-poor population.\\  $\feh = -1.7 \pm 0.1$, age $13 \pm 1$ Gyr.

\item[Population B:]  Sgr metal-poor population.\\  $\feh = -1.2 \pm 0.1$, age $11 \pm 1$ Gyr.

\item[Population C:]  Sgr intermediate population.  \\ Three sub-populations:\\ C1: $\feh = -0.65 \pm 0.1$, age $7 \pm 0.6$ Gyr,\\ 
C2: $\feh = -0.45 \pm 0.1$, age $6 \pm 0.6$ Gyr,\\ C3: $\feh = -0.3 \pm 0.1$, age $5 \pm 0.6$ Gyr.

\item[Population D:]  Sgr young population.\\  $\feh = -0.05 \pm 0.15$, age $2.5 \pm 0.5$ Gyr.

\item[Population E:]  Sgr very young population.\\  $\feh = +0.55 \pm 0.05$, age $0.75 \pm 0.1$ Gyr.

\end{description}


Younger stellar populations might be realistically assumed to be more tightly bound than older populations, which have had more time to diffuse outward in the dwarf.
Motivated by the presence of such population gradients within other Galactic dSphs (e.g., Tolstoy et al. 2004)
for which the more metal-rich populations are more centrally concentrated and have lower velocity dispersion,
we tag particles in our best-fit model of the Sgr dwarf to correspond to different stellar populations according to their total energy in the pre-interaction dSph.
The energy of an individual particle in the satellite is given by
\begin{equation}
E[i] = \frac{1}{2} v[i]^2 + \Phi[i]
\end{equation}
where $v[i]$ is the velocity and $\Phi[i]$ the internal potential energy of particle $i$ relative to the satellite.  
For convenience, we define $E^{\ast}$ to be an order-sorted version of the energy distribution,
ranging from 0 (most strongly bound particle) to 1 (least bound particle) with a uniform distribution in the range $0 - 1$.

\begin{figure}
\plotone{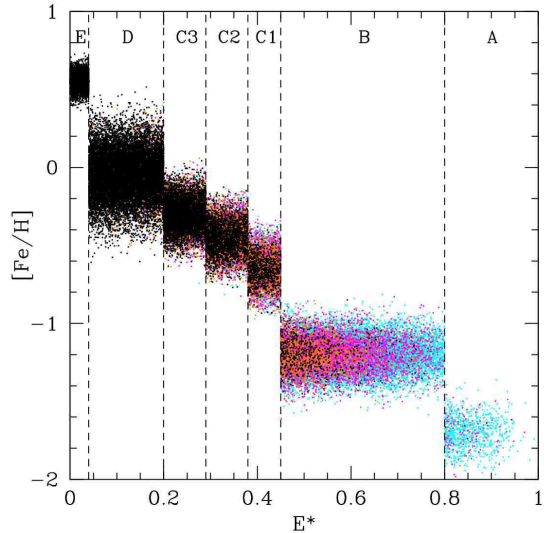}
\caption{Metallicity assigned to all particles in the simulation as a function of their initial energy $E^{\ast}$ in the original satellite.  
Dashed lines indicate the adopted energy divisions between
the populations, as named in the list in Section \ref{enrichmodel.sec}.
Colors of points indicate
the pericentric passage on which the particle becomes stripped from the satellite (as in Figure \ref{XYgrid.fig}).}
\label{fehassign.fig}
\end{figure}

We assign particles with a given range of $E^{\ast}$ to each population as illustrated in Figure \ref{fehassign.fig}, the locations of the energy divisions
(and hence the total number of particles assigned to each burst of star formation) are free parameters.  
Once a particle has been assigned to a given population, we assign it a metallicity
and age based on a Gaussian random selection about the mean and standard deviation of the population given in the list above, and randomly choose an
infrared color uniformly from the
range $J-K_s = 1.0 - 1.1$.
Adopting Marigo et al. (2008) isochrones\footnote{We use the interface made available by L. Girardi at
http://stev.oapd.inaf.it/cgi-bin/cmd}
 with a Chabrier (2001) IMF for each of the simulated populations, we calculate the absolute $K_s$ magnitude for each simulated particle, adopting a 
 50\%/50\% mix of RGB and AGB stars.
This particular mix of RGB/AGB stars is not intended to match any particular distribution, but to illustrate the general range of magnitudes expected 
for stars in each evolutionary phase.
We note that not all of the Siegel et al. (2007) populations will produce M-giants: Population A in particular is too old and 
metal-poor.\footnote{Indeed (as discussed by Siegel et al. 2007), the M54 metal-poor population may have formed separately from Sgr and later been accreted by the dwarf,
since M54 has a distinct stellar population and radial profile and Sgr is nucleated even ignoring this population (e.g., Monaco et al. 2005).}
All this means, however, is that we would not expect regions of the stellar streams populated by these stars to be traced by M-giants
of those metallicities, and we therefore do not calculate $K_s$ magnitudes for these particles.

Strictly speaking, the enrichment model presented here is unphysical since
it assumes that {\it all} of the stellar populations were present in Sgr at all times during its $\sim 8$ Gyr interaction history with the Milky Way, despite
the fact that some of these populations didn't form until many Gyr later.
This is an inescapable consequence of the $N$-body method adopted in this paper --- particles cannot readily be introduced into the simulation at arbitrary times.
Ideally, a fully hydrodynamical simulation could build a gas-phase component into the model dwarf, and permit this to form stars at the appropriate times.
How to retain an appreciable gas supply in Sgr until only a Gyr ago is a challenging question in its own right, but claims of a detection of gas
removed from Sgr as recently as $\sim 0.4$ Gyr ago by its most recent passage through the extended Galactic disk have been made by Putman et al. (2002).
Since the $N$-body particles only represent dynamical {\it tracers} however, we content ourselves with noting that no appreciable 
quantities of any given stellar population are tidally stripped from
the dwarf in this model before they are thought to have formed (see horizontal dashed lines in Figure \ref{TubEstar.fig}).


\subsection{Matching the Observed MDF}
\label{matchingMDF.sec}

\begin{figure}
\plotone{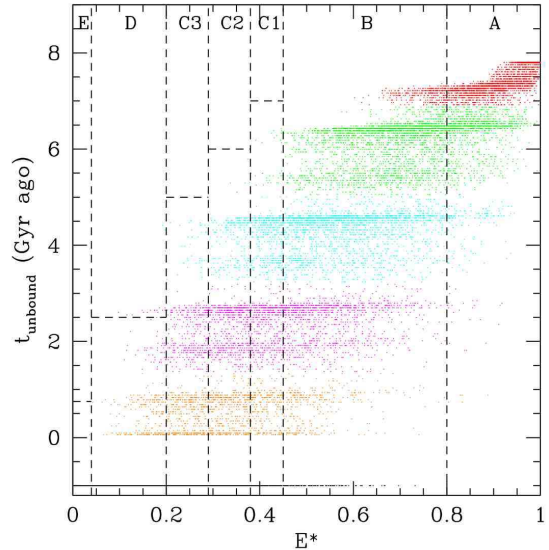}
\caption{\scriptsize Time a given particle is stripped from the model dSph as a function of initial energy $E^{\ast}$ (values of $t_{\rm unbound} = -1$ indicate
particles bound to the dSph at the present day).  
Colors correspond to specific of orbits in the lifetime of the dwarf (see Fig. \ref{OrbitPlot.fig} for key).
Vertical dashed lines indicate the divisions assigned between stellar populations (as in Figure \ref{fehassign.fig}), horizontal dashed lines
indicate the formation time of each population.  Note that while 
mass loss is always occurring, the majority (i.e., corresponding to the highest point density)
is concentrated into relatively brief periods surrounding pericentric passages.}
\label{TubEstar.fig}
\end{figure}

Tidal stripping of a satellite on an eccentric orbit such as that of Sgr occurs nearly impulsively, 
with the majority of tidal debris stripped in the brief time interval that the satellite is passing
through the pericenter of its orbit.  
Although the overall trend with successive pericentric passages is to remove stars at successively smaller radii or energies
(i.e., `onion-skin' style stripping models, see discussion by Mart{\'{\i}}nez-Delgado et al. 2004), 
each mass loss event does not impose a sharp tidal cutoff on the stellar populations of the dwarf.
Rather, the rapidly changing tidal radius, the impulsive nature of the stripping, and the dependence of 
the effectiveness of tidal stripping on the orientation of individual stellar orbits
(see, e.g., Read et al. 2006a)
means that on any given pericentric 
passage Galactic tides reach deeply into the satellite and can
unbind stars from a variety of populations.
This does not completely evacuate the energy/radius range of particles in the  satellite,
and the remaining stars rapidly relax back to a Plummer model over a dynamical time ($\sim 0.1 - 0.2$ Gyr; see, e.g., discussion by Pe{\~n}arrubia et al. 2008, 2009).
Despite the sharp divisions that we introduced in the distribution of stellar populations within the model satellite, there is
significant overlap in the populations expelled
from the satellite into the tidal tails over the lifetime of the dwarf.
This is illustrated in Figure \ref{TubEstar.fig}: While particles assigned to Population A are almost entirely stripped from the dwarf
during the first two to three orbital periods, stars in Population B 
are present in many successive epochs of tidal debris and still have a residual bound population (black points
at $t_{\rm unbound} = -1$) in the dwarf at the present day.
Stars in Population D are present in only the most recent tidal debris.
The trend of stellar populations along the tidal streams is further affected by the angular phase smearing caused by the differential orbital velocities
of stripped stars from each epoch of tidal debris.

\begin{figure*}
\epsscale{0.8}
\plotone{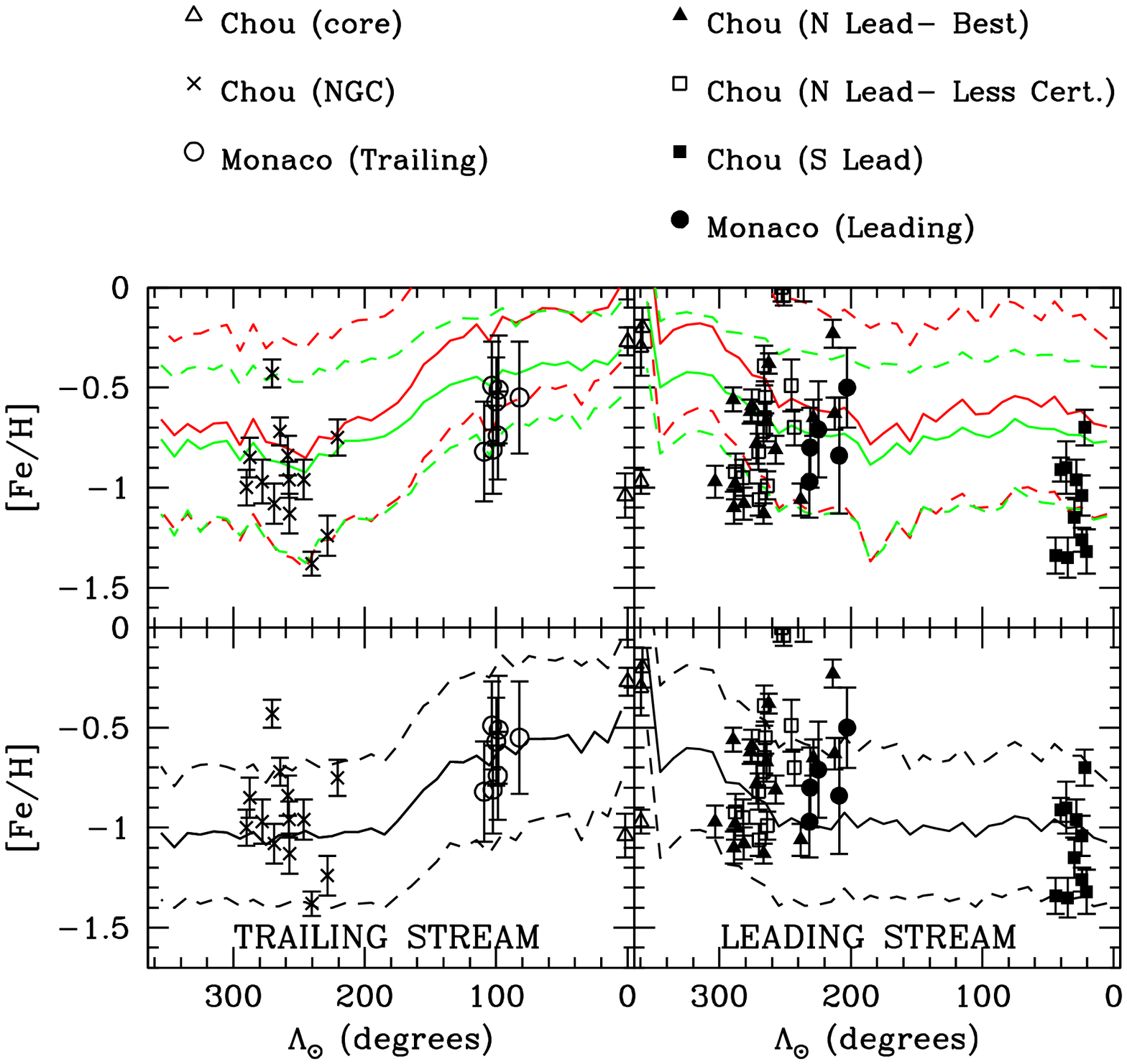}
\caption{\scriptsize Metallicity as a function of orbital longitude along the leading/trailing (right/left panels respectively)
Sgr tidal streams for the Chou et al. (2007) and Monaco et al. (2007) observational samples.  Different symbols are used to indicate stars in the various leading/trailing
arms subsamples described by these authors.
In the middle and lower panels we overplot the mean metallicity (solid lines) for $N$-body particles in the most recent 5 Gyr of tidal debris.  The range of typical 
metallicities about this average is indicated by the dashed lines.  The red/green lines respectively in the middle panels
indicate two methods of debris-tagging which do not result in a trend matching the observational data, while the black lines in the lower panel generally
reproduce the observational trends (see discussion in \S \ref{matchingMDF.sec}).}
\label{MetalObs_FeH.fig}
\end{figure*}

By adjusting the relative sizes of each of the stellar populations (i.e., sliding the vertical lines in 
Figures \ref{fehassign.fig} and \ref{TubEstar.fig}) it is possible to adjust the relative balance
of stars of a given metallicity present in each debris era (i.e., point `color'), and thereby as a function of orbital longitude along the Sgr tidal streams.
In Figure \ref{MetalObs_FeH.fig}  we plot the metallicity of stars observed by Chou et al. (2007) and Monaco et al. (2007; their SARG subsample) 
as a function of orbital longitude, overplotted with the mean metallicity of the simulated Sgr stream using three different assumptions about
the energy ranges $E^{\ast}$ assigned to each of the Populations A-E.  
The simplest possible assignment
in which Populations A/B/C/D/E correspond to uniformly wide energy ranges 
$E^{\ast} = 1.0-0.8 / 0.8-0.6 / 0.6-0.4 / 0.4-0.2/ 0.2-0.0$ respectively fails to match the observational data.  As illustrated by the red curves in
Figure \ref{MetalObs_FeH.fig}, the highest-metallicity debris is over-represented in the regions of the stream nearest the satellite
(i.e., $\Lambda_{\odot} \approx 250^{\circ}$ and $100^{\circ}$ in the leading/trailing  streams respectively), and Population C is over-represented
relative to Populations A and B at larger angular separations (i.e., the model stream is too metal-rich at $\Lambda_{\odot} \approx 30^{\circ}$ 
and $250^{\circ}$ in the leading/trailing streams respectively).
 It is possible to suppress the presence of the youngest stellar populations in the Sgr stream
by reducing the range in $E^{\ast}$ assigned to them, and the green curves in Figure \ref{MetalObs_FeH.fig} show the result of adopting
energy ranges $E^{\ast} = 1.0-0.8 / 0.8-0.6 / 0.6-0.2 / 0.2-0.04/ 0.04-0.0$ for Populations A-E respectively.
While this improves the agreement in the regions of the tidal tails nearest Sgr, the mean model metallicity is still too high at larger angular separations.

We find that a reasonable fit to both the low- and high-angular separation observational data is achieved with a choice of energy ranges
$E^{\ast} = 1.0-0.8 / 0.8-0.45 / 0.45-0.38 / 0.38-0.29 / 0.29-0.20 / 0.20-0.04/ 0.04-0.00$ for Populations A/B/C1/C2/C3/D/E respectively (black lines in Figure
\ref{MetalObs_FeH.fig}).
Given the large spread of the observational measurements and the large number of assumptions made by our debris-tagging method, a more precise
determination of the best-fit choice of energy ranges is not warranted.  Rather, we conclude that for some choice of the energy ranges it is possible to broadly
fit the observational trends: The mean metallicity of the model Sgr stream decreases from \feh $\sim -0.4$ in the core to \feh $\sim -1.0$  at large angular separations.

\begin{figure*}
\plotone{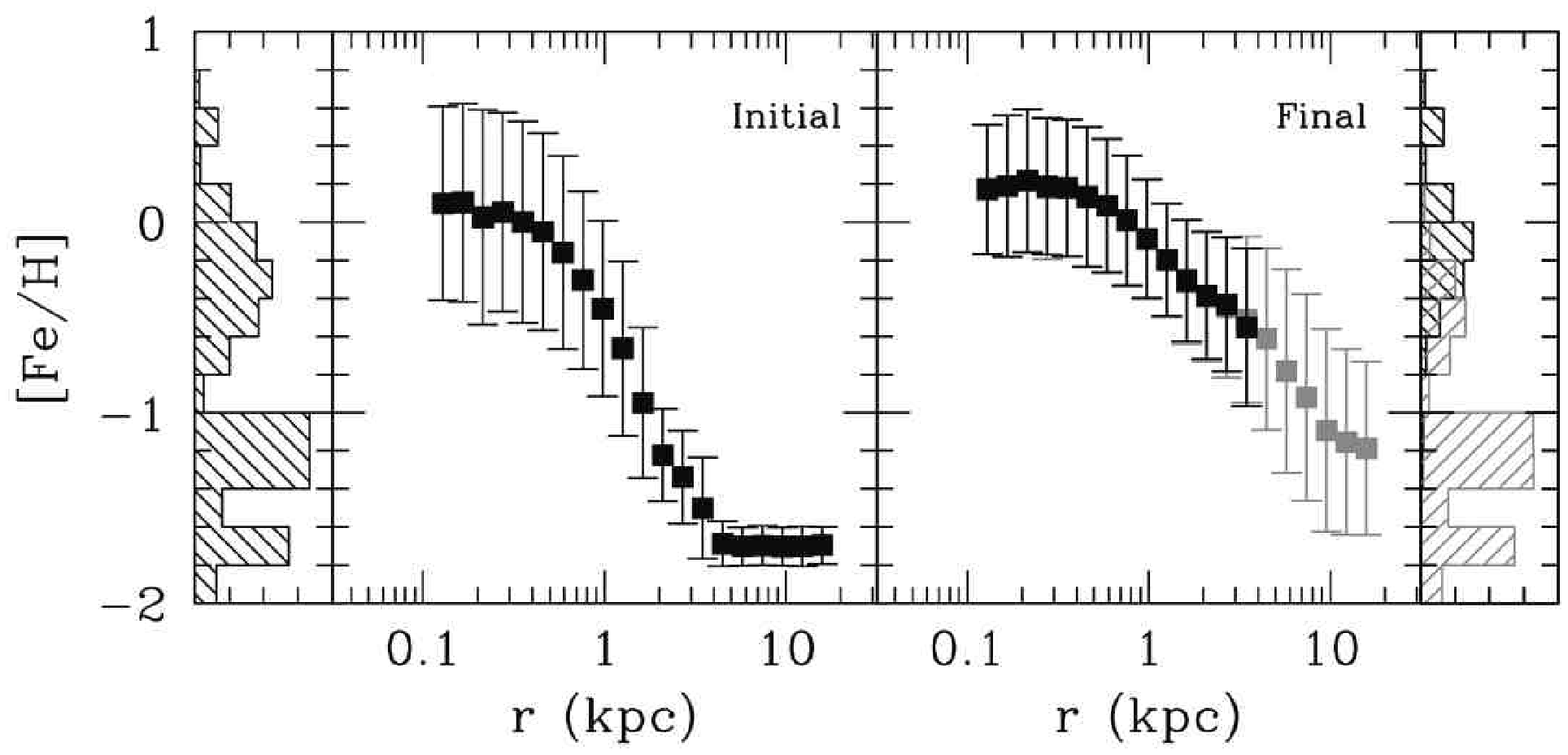}
\caption{Metallicity as a function of radius in the pre-interaction (left-hand panel) and present day (right-hand panel) Sgr dSph.  
Filled squares represent the mean value of \textrm{[Fe/H]}
at a given radius, the vertical bars represent the $1\sigma$ spread about this mean value at each radius.  Black points represent  \textrm{[Fe/H]} for particles
bound to the Sgr dwarf, grey points represent unbound stars in the tidal streams.  Inset histograms at the left and right edges represent the MDF  of the intial
and present-day Sgr dwarf respectively: Note the difference between the present-day MDF of bound particles (black histogram) and particles in the tidal
streams (grey histogram).}
\label{rfeh.fig}
\end{figure*}

The above choice of energy ranges corresponds to a reasonably strong $\sim 2$ dex metallicity
gradient with radius in the initial model Sgr dSph (Figure \ref{rfeh.fig}, left-hand panel), consistent
with the prediction of Chou et al. (2007), and slightly stronger than observed in the Sculptor dSph by Tolstoy et al. (2004).
The strong initial gradient is largely erased by the process of tidal stripping, leaving a relatively weak  metallicity gradient $\sim 0.8$ dex
remaining by the present-day  (Figure \ref{rfeh.fig}, right-hand panel).
The MDF of the bound core of the model Sgr dwarf (Figure \ref{rfeh.fig}, right-inset black histogram) is therefore significantly different
than both the MDF of the original dSph (Figure \ref{rfeh.fig}, left-inset black histogram) and the MDF of the stellar tidal debris that Sgr has
contributed to the Galactic halo (Figure \ref{rfeh.fig}, right-inset grey histogram).
If such strong tidal evolution of the MDF is common for Galactic satellites, it may partially explain the observed chemical abundance mismatch 
between Galactic halo stars and the present-day dSph population (e.g., Geisler et al. 2005; and references therein).
Rather than requiring the early accretion onto the Milky Way of metal-poor, high \textrm{[$\alpha$/Fe]} satellites that are completely disrupted
by the present day (e.g., Font et al. 2006), the metal-poor halo stars may simply have originated in the dSphs with which we are familiar, and in which
few traces of the originally sizeable metal-poor populations now remain (e.g.,  Majewski et al. 2002; Mu{\~n}oz  et al. 2006; Chou et al. 2009).


\subsection{K magnitudes of M giants}
\label{mgiant_kmags.sec}

One of the challenges in attempting to trace Sgr debris in the 2MASS M-giant view originally provided by Majewski et al. (2003)
was that as the metallicity of stars evolves along the stream, so does the zeropoint of the photometric distance calibration.  Since the distance calibration was tied
to the relatively metal-rich Sgr core, distances become systematically more inaccurate with increasing separation from the core.
Now that we have a model for the evolution of metallicity along the streams, we can perform the comparison of simulations and observational data in a purely
observational apparent $K_s$ magnitude space.

\begin{figure*}
\epsscale{0.8}
\plotone{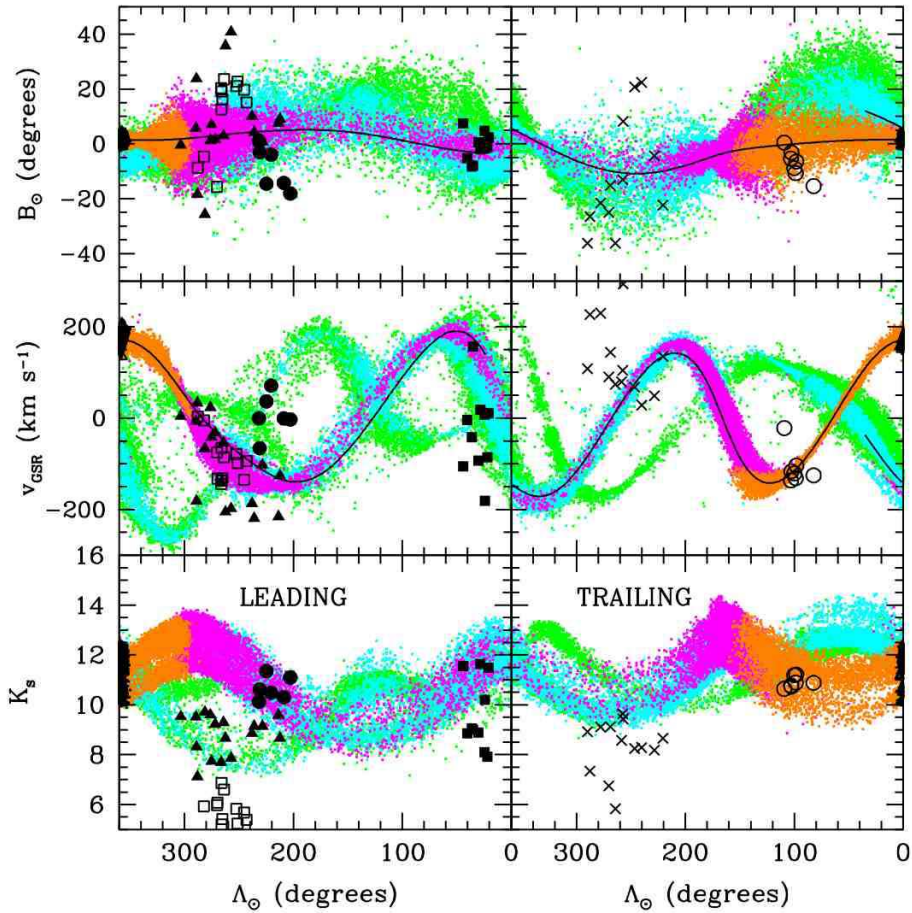}
\caption{\scriptsize $N$-body debris particles (colored points) for the best-fit model, 
overlaid with stars (see key in Figure \ref{MetalObs_FeH.fig}) for which metallicity information has been acquired by Chou et al. (2007) and Monaco et al. (2007).
Note that for purposes of comparison we plot more epochs of tidal debris here than shown in Figure \ref{ModelSummary.fig}.
For clarity we omit the Sgr core sample of Chou et al. (2007; i.e., open triangles) from this figure since they would be indistinguishable from the black-colored $N$-body debris points.  The upper/lower `branches' of the model stream at slightly different $K_s$ magnitudes represent the RGB and AGB evolutionary phases respectively.
The solid line in the upper and middle panels indicates the orbital path of the Sgr dwarf core.}
\label{MetalObs.fig}
\end{figure*}

In Figure \ref{MetalObs.fig} we plot the apparent $K_s$ magnitude, angular position $(\Lambda_{\odot}, B_{\odot})$, and radial velocity
$v_{\rm GSR}$ for the simulated Sgr dwarf, overlaid with the Chou et al. (2007) and Monaco et al. (2007) data.
We note that the $N$-body model generally succeeds in reproducing the radial velocity trends indicated by these data, although the
Chou et al. (2007) trailing/leading arm sample of stars wrapped into the North/South Galactic Hemisphere (crosses/filled squares 
in Figure \ref{MetalObs.fig}) may constitute evidence for a more massive Sgr progenitor in order to produce
extended wraps of the stream that are broader in radial velocity space.  The $N$-body models also suggest that some of the observational data
may originate elsewhere than the Sgr stream: A few of the crosses, three-five filled circles, and two open circles in Figure \ref{MetalObs.fig} have radial
velocities strongly inconsistent both with the other data points and the $N$-body models.

\begin{figure*}
\epsscale{0.8}
\plotone{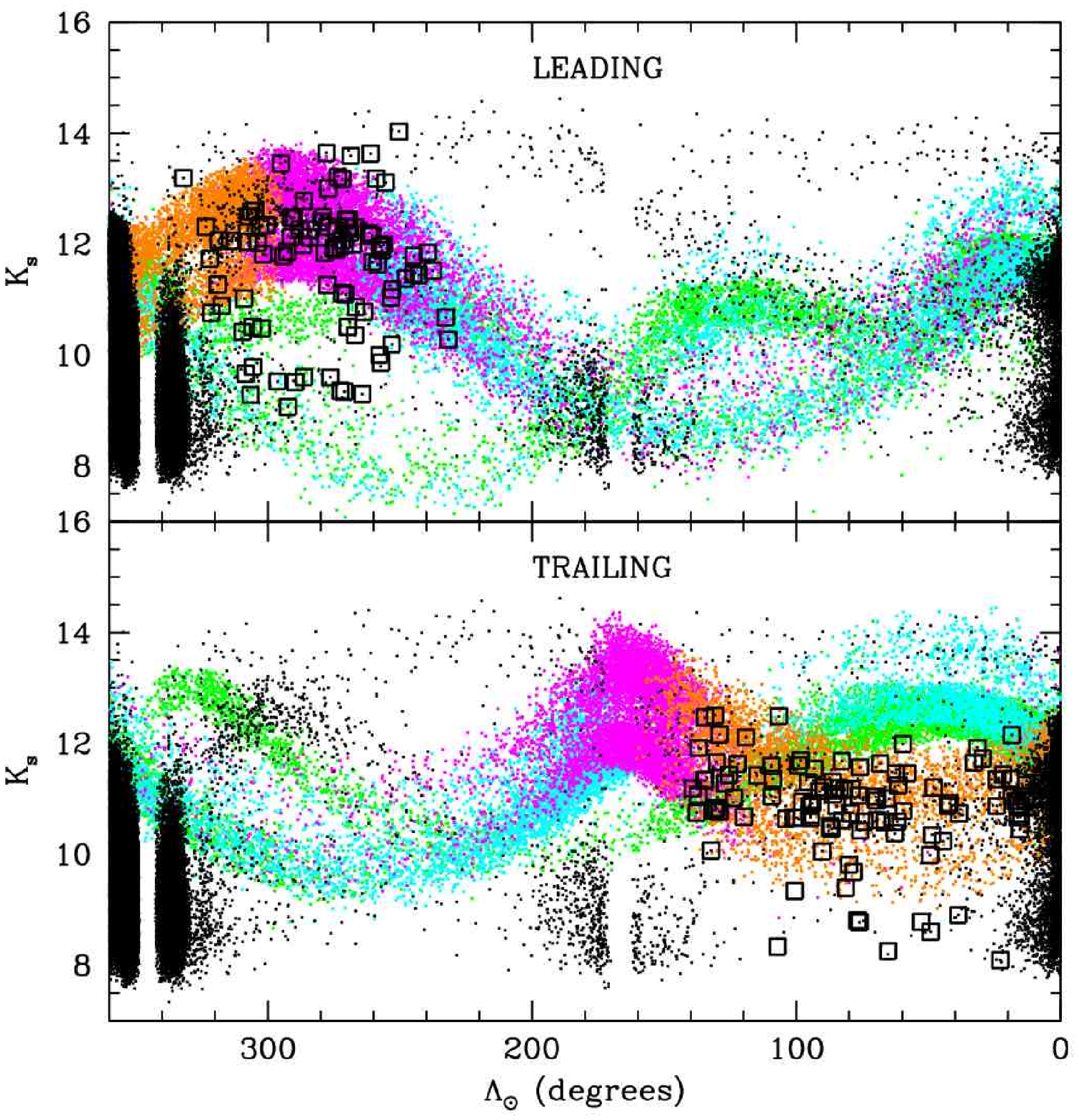}
\caption{\scriptsize Apparent $K_s$ magnitude as a function of orbital longitude for leading and trailing tidal arms.  
Colored points represent debris from the $N$-body model (for 
purposes of comparison we plot more epochs of tidal debris here than shown in Figure \ref{ModelSummary.fig}), overlaid with
small black points representing M-giants from the sample of Majewski et al. (2003; see their Figure 8).  
We also overlay the subsample of M-giants with radial velocities (open boxes) pruned in Figure \ref{rvcull.fig}.}
\label{LamKapp.fig}
\end{figure*}

The agreement between the apparent $K_s$ magnitudes of the $N$-body model and the Monaco et al. (2007) data appears to be relatively good.
Unfortunately it is difficult to evaluate this agreement for the Chou et al. (2007) data since these authors focussed on the brightest available M-giants,
which are expected to be biased several or more $\sigma$ brighter than the mean for the stream (see discussion by Chou et al. 2009).
A more instructive comparison is to the M-giant survey of Majewski et al. (2003).
In Figure \ref{LamKapp.fig} we plot the apparent $K_s$ magnitude for the simulated Sgr dwarf, overlaid with the original M-giant sample from Majewski et al. (2003; i.e.,
similar to their Fig. 8) and the subsample with radial velocity data which we culled from Law et al. (2004) and Majewski et al. (2004)
in Figure \ref{rvcull.fig} (open boxes in Figure \ref{LamKapp.fig}).  The $K_s$ magnitudes derived from the  model
match the larger M-giant sample of the dynamically young sections of the tidal streams (i.e., $\Lambda_{\odot} \lesssim 150^{\circ}$ for the trailing arm, and
$\Lambda_{\odot} \gtrsim 220^{\circ}$ for the leading arm) well: Over 95\% of the radial velocity subsample have $K_s$ magnitudes consistent with
the $N$-body model.  The remaining 5\% may be Galactic interlopers at smaller distances whose colors and radial velocities happened to be coincident
with those of the Sgr stream, although a few located around $(\Lambda_{\odot}, K_s) = (280^{\circ}, 9.5)$ may be associated with the T1 wrap of the trailing arm stream
instead of the fainter (and more distant) L1 stream since the radial velocity trends of the L1 and T1 wraps are expected to cross in this region of the sky.

There are a few regions of stars in Figure \ref{LamKapp.fig} that are not explained by the model however.
In particular there is a distribution of M-giants at $K_s \sim 13-14$ at all orbital longitudes that are not reproduced by the $N$-body model.
At $\Lambda_{\odot} \sim 180^{\circ} - 200^{\circ}$ these faint M-giants arguably appear to join smoothly to the trailing arm trend
(see also Newberg et al. 2003), suggesting
(Pakzad et al. 2004) that
the trailing arm may {\it not} turn around at this location to approach closer to the Sun again (as the $N$-body model does) but instead forms
a wrap around the entire sky at relatively constant $K_s \sim 13-14$.
Such a wrap is strongly inconsistent with the predictions of all $N$-body models
to date and, if genuine, may be evidence for a strong evolution in the orbital path of Sgr $\sim 3$ Gyr ago.

\subsection{Caveats}

\subsubsection{Relation to Other Tracers}

The picture presented here of the evolution in the mean metallicity along the Sgr tidal arms is  biased towards more metal-rich stars, since it has been fitted
to observations of intrinsically metal-rich M-giant tracers.  Other tracers such as blue horizontal branch (BHB) stars (e.g., Yanny et al. 2009), 
RR Lyrae (RRL; Vivas et al. 2005, Prior et al. 2009), or K-giants 
(Starkenburg et al. 2009; see also  Dohm-Palmer et al. 2001)
might reasonably be expected to differ from the trends calculated here.
It is unclear, however, how stars more metal-poor than $\feh \sim -2$ (as detected by Yanny et al. 2009) might arise in 
large numbers in relatively young sections of the Sgr stream (i.e., the L1 arm at $\Lambda_{\odot} \sim 250^{\circ}$) since no remnants of populations so metal
poor remain in the core of Sgr today (Siegel et al. 2007).
If this metal-poor stellar population has indeed recently been completely stripped from Sgr it may provide a useful constraint upon future
models of the distribution of stellar populations within the initial Sgr dSph.
We discuss this issue further in \S \ref{otherobs.sec}.

\subsubsection{Uniqueness}

The fact that we have been able to find a method for tagging particles in the 
initial satellite to correspond to the stellar populations seen in the tidal streams should not be taken to mean
that this structure of the initial satellite is uniquely correct, that the optimal ranges in $E^{\ast}$ found for each star formation episode A-E correspond exactly to the total
stellar mass produced in the bursts, or even that the method is a particularly good description of the tidal stripping of specific stellar
populations within a real Galactic dSph.  
The mass loss history of the satellite is a strong function of the adopted model, and different values for the total mass, scale length,
or interaction time of the
dwarf will produce different amounts of overlap between 
various eras of tidal debris, and various ranges in orbital energy within the original dwarf.  Similarly, if we were to adopt a different physical
model for Sgr (e.g., a cuspy Hernquist profile, or a more 
physically motivated two-component King-profile stellar core combined with an extended NFW dark matter halo), 
or consider the possible conversion of an originally disky dwarf to a dSph via tidal stirring (e.g., Mayer et al. 2006)
the mass loss history could also be very different (see, e.g., Read et al. 2006ab).  
The specific distribution and mass normalization of individual stellar populations is therefore contingent on the adopted model and should not be interpreted as
a general result, but can be refined as more observational data become available.




\section{Discussion: Relation to Possible Detections of Sgr Tidal Debris}
\label{otherobs.sec}

The model presented in this paper has the attractive quality that it provides a reasonable match to the major dynamical 
characteristics of Sgr tidal debris  measured from wide-field surveys.
It is worthwhile also investigating the degree of correspondence between
the $N$-body model and various narrow-field observations of subgiant stars, K-giants, and RRL
that have been postulated to trace the Sgr tidal stream.  

\begin{figure*}
\epsscale{0.8}
\plotone{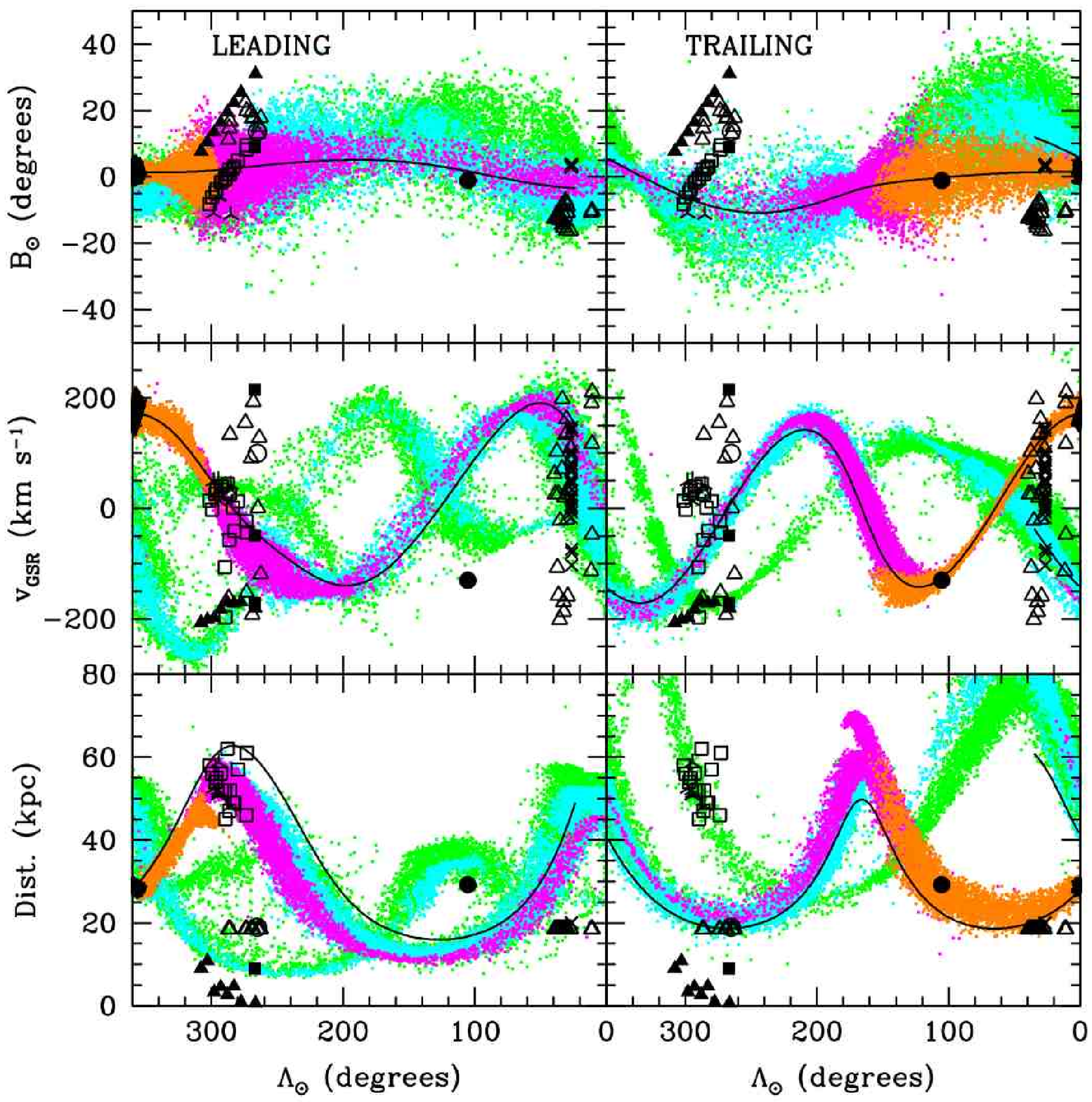}
\caption{\scriptsize Model (colored points) of the Sgr stream overlaid with various detections of potential Sgr tidal debris.
We include 2 Gyr more of  tidal debris in this plot (green points) than in previous plots (e.g., Figure \ref{ModelSummary.fig}) to indicate the possible path of debris
if Sgr has been orbiting the Milky Way for more than $\sim 5$ Gyr.  Overlaid on the $N$-body model
is observational data from Vivas et al. (2005; open boxes), the SDSS `Stripe-82' (Cole et al. 2008;
Watkins et al. 2009; filled circle), Duffau et al. (2006; open circle), `VOD' observations from
Vivas et al. (2008; filled boxes), K-giant stars from Starkenburg et al. (2009; skeletal triangles) and Kundu et al. (2002; filled triangles),
RR-Lyrae stars from Prior et al. (2009; open triangles), and red clump stars from Majewski et al. (1999; crosses).  Note that 
at $\Lambda_{\odot} \approx 260^{\circ} - 300^{\circ}$ it is difficult to distinguish the open circle
of Duffau et al. (2006; because it lies atop the open triangles of Prior et al. 2009 in
$\Lambda_{\odot}$ vs distance space), and the skeletal triangles of Starkenburg et al. (2009; because they
lie amidst the open boxes of Vivas et al. 2005).}
\label{OtherObs.fig}
\end{figure*}

In Figure \ref{OtherObs.fig} we plot the $N$-body model of the Sgr stream overlaid by a selection of detections from recent
narrow-field surveys (this comparison is not exhaustive, see also discussion by Majewski et al. 2003).
We discuss each data set in detail below.
Note that we do not consider possible associations with globular clusters or faint dwarf satellites; these are treated
separately in a forthcoming contribution (Law et al. {\it in prep.}).

\subsection{SDSS Stripe 82}

Cole et al. (2008) used a maximum likelihood method to characterize the spatial properties of F-turnoff stars in stripe 82 of the SDSS.
They observed tidal debris located at $(\alpha, \delta, R) = (31.37^{\circ} \pm 0.26^{\circ}, 0.0^{\circ}, 29.22 \pm 0.20 \, {\rm kpc})$,
corresponding to Sgr longitudinal coordinates $(\Lambda_{\odot}, B_{\odot}) = (105.2^{\circ}, -1.15^{\circ})$.
As illustrated by Figure \ref{OtherObs.fig} (filled circles) the T1 wrap of the $N$-body model is an extremely good match to these 
F-turnoff stars in both angular position and distance
(matching to within $0.7^{\circ}$ in $B_{\odot}$, and 1 kpc in distance).
Cole et al. (2008) measured the width of the stream perpendicular to the line of sight at this $\Lambda_{\odot}$ to have a FWHM $ = 6.74 \pm 0.06$ kpc.
The $N$-body stream is slightly fatter with FWHM $ = 7.9$ kpc, but given the single component mass-follows-light model 
employed for Sgr the agreement with the apparent
width of a specific stellar population is extremely good.

The radial velocity of a sample of RRL in Stripe 82 has recently been measured by Watkins et al. (2009), who detect a
component with $v_{\rm GSR} = -130$ \kms, which is matched by the mean radial velocity of the T1 stream at this position
to within 5 \kms.  Watkins et al. (2009) also measure the metallicity of these  RRL, finding that 
$\langle  \feh \rangle = -1.41 \pm 0.19$.  This value is appreciably more metal-poor than either the M-giant sample or
the mean of the $N$-body model at this longitude ($\feh \sim -0.7$).  This difference in measured metallicity emphasizes the strong sensitivity
of derived metallicities of a stellar stream to the biases imposed by the type of stellar tracer used (e.g., metal-poor populations do not generally
form M giants while metal-rich populations do not generally form RRL).
Thus, until an unbiased assessment of the MDF along the Sgr stream is obtained, models such as that presented in \S \ref{feh.sec},
which was constrained by a single type of metallicity-biased population tracer, must be considered as merely a demonstration of how
MDF-variations can originate.

\subsection{RRL and K-giants at $\Lambda_{\odot} \sim 270^{\circ} - 300^{\circ}$}

Vivas et al. (2005) published VLT spectroscopy of 16 RRL from the QUEST survey
in the range $\Lambda_{\odot} \sim 270^{\circ} - 300^{\circ}$.  All but 1-2 of these stars,
(which have highly discrepant velocities, at  $v_{\rm GSR} = -106$ and $-197$ \kms) are in agreement 
 with the angular position, distance
and radial velocity trend of the L1 arm of the $N$-body model (see Figure \ref{OtherObs.fig}, open boxes).
Starkenburg et al. (2009) have also detected an overdensity of 5 K-giant stars at similar orbital longitude
that are well-matched to the distance and radial velocity of the $N$-body model stream (Figure \ref{OtherObs.fig}, skeletal
triangles), although they appear to be located towards 
the edge of the stream in $B_{\odot}$.\footnote{Note that the star added to the Starkenburg et al. (2009) group using the `4dist' $\leq 0.08$ criterion
is not as well-matched to the $N$-body stream as the other 5 stars, with the former discrepant by $\sim 60$ \kms.}

Both samples of stars are well-reproduced by the $N$-body model, although we note that the Vivas et al. (2005)
and Starkenburg et al. (2009) samples have mean metallicity $\langle  \feh \rangle = -1.77$ and $-1.68$ respectively.
These metallicities are entirely consistent with each other, but significantly
more metal-poor than the M-giant sample, which has $\langle  \feh \rangle \sim -0.9$.
As was the case for SDSS Stripe 82, this reinforces the conclusion that stellar populations significantly more metal-poor than
the M-giant subsample exist at a range of orbital longitudes throughout the Sgr tidal streams.  As expected on the basis
of dynamical age, the RRL of Vivas et al. (2005) which correspond to  tidal debris $\sim 1.5 - 3$ Gyr old
(i.e., magenta points in Figure \ref{OtherObs.fig}) are slightly more metal-poor than the RRL in SDSS Stripe 82 which correspond
to $\sim 0 - 1.5$ Gyr old debris.

We note that an additional sample of metal-poor K-giant stars in a similar longitude range was also pointed out by Kundu et al. (2002; filled triangles
in Figure \ref{OtherObs.fig}).  While the L2/T2 wraps of the $N$-body stream match the radial velocity trend of these stars, the angular positions are discrepant by
up to $20^{\circ}$ for L2 and $30^{\circ}$ for T2.  At a typical distance $\sim 5$ kpc the Kundu et al. (2002) stars are also much closer than expected for the T2 arm
($d \sim 50$ kpc), but may be consistent with the L2 arm ($d \sim 10$ kpc).  If these stars genuinely belong to the Sgr stream, it is most likely that they belong to the L2
wrap, and may suggest that the angular coordinates and/or distance of this wrap in the $N$-body model require adjustment.

\subsection{Virgo: The VSS and the VOD}

Recent years have witnessed a wealth of substructure discovered in the direction of Virgo, which we
follow Mart{\'{\i}}nez-Delgado et al. (2007) in classifying into the Virgo Stellar Stream (VSS)
and the Virgo Overdensity (VOD).

The VSS (as traced by QUEST RRL; Duffau et al. 2006) is a relatively well-defined feature that lies 
at an angular position $(\Lambda_{\odot},B_{\odot}) = (265^{\circ}, 14^{\circ})$, i.e., projected
on the outer edge of the L1/L2 streams and significantly off the T1/T2 streams in sky position.
At a distance of 19 kpc, it is roughly coincident with the L2/T1 stream, but has a radial velocity ($v_{\rm GSR} = 99.8$ \kms; Duffau et al. 2006)
that is discrepant with these wraps by 220/94 \kms respectively.  Curiously, the proper motion of the VSS
($[\mu_{\alpha} \textrm{cos} \delta, \mu_{\delta}] = [-4.85 \pm 0.85, 0.28 \pm 0.85]$; Casetti-Dinescu et al. 2009) is reproduced by the T1 stream
to within $1\sigma$ ($[\mu_{\alpha} \textrm{cos} \delta, \mu_{\delta}] = [-4.18 \pm 0.41, 0.04 \pm 0.25]$), but given the angular position and radial
velocity mismatch we conclude that the VSS is unlikely to be associated with the Sgr stream.

In contrast, the VOD (Juri{\'c} et al. 2008) is a much less well-defined, diffuse clump of stars around $(l,b) = (300^{\circ}, 65^{\circ})$
that covers more than $\sim 1000$ deg$^2$ on the sky.
The possible association of the VOD with Sgr has been addressed in detail by Mart{\'{\i}}nez-Delgado et al. (2007), who noted
that it was roughly coincident with tidal debris from models of the Sgr dwarf (LJM05) disrupting in an axisymmetric, oblate Galactic halo.

In the triaxial halo model derived in this contribution, we find that the model Sgr stream arcs significantly over the solar
neighborhood, and tidal debris in the L1 tidal stream at the angular coordinates of the VOD lies much too far away
($\sim 46$ kpc for the Sgr stream versus $\sim 5-17$ kpc for the VOD) to be associated  (see Figure \ref{OtherObs.fig},
filled squares).  Similarly, both the T1 and T2
wraps lie too far away to produce the VOD, in addition to 
lying in the opposite angular direction $B_{\odot}$.

Interestingly, this angular position and distance {\it does} correspond closely to the predicted location of the secondary L2
wrap of leading Sgr tidal debris, suggesting that the VOD may be evidence for old tidal debris torn from Sgr $\gtrsim 3$ Gyr ago.
If such an identification is correct, we should expect the radial velocity of the VOD to be $v_{\rm GSR} = -131 \pm 22$ \kms.
Recent observational work on QUEST RRL in the direction of the VOD  by Vivas et al. (2008) derived radial velocity signatures
at $v_{\rm GSR} = +215, -49, -171$ \kms.  The first (and by far most significant) two of these velocity structures are strongly discrepant
with the expected signal by $\sim 345$ and $\sim 80$ \kms respectively.  The third structure is possibly consistent with the velocity
signature expected of the L2 Sgr stream, differing by only $\sim 40$ \kms in this part of the simulated debris where the $N$-body stream is ill-constrained.
Only 7\% of the QUEST RRL (i.e., 3/43) are contained in this $v_{\rm GSR} = -171$ \kms velocity structure however,
and we therefore concur with the conclusion of Vivas et al. (2008) that while Sgr may make some minor contribution to the VOD
it is not the dominant origin of the stellar overdensity.

\subsection{SEKBO RRL}

An additional sample of RRL stars from the SEKBO (Moody et al. 2003) survey has recently been studied by
Keller et al. (2008), with radial velocities presented by Prior et al. (2009).  The Prior et al. (2009) subsample
focuses on two groups of RRL which they group into the ``VSS region'' at $\alpha = 12.4 - 14$ hr (i.e., 
$\Lambda_{\odot} \approx 260^{\circ}-290^{\circ}$) and the ``Sgr region'' at $\alpha = 20 - 21.5$ hr
(i.e., $\Lambda_{\odot} \approx 10^{\circ}-40^{\circ}$).  These two groups of stars are plotted against our $N$-body
model in Figure \ref{OtherObs.fig} (open triangles).

Considering first the angular coordinate plot $\Lambda_{\odot}$ vs. $B_{\odot}$, we note that neither of these groups
of stars are well-centered on the predicted path of the $N$-body tidal stream, lying at the extreme outer edge of the L1/L2 streams
and fully outside the angular path of the T1/T2 streams.  This is consistent with Figure 15 of Keller et al. (2008): The plane
of the SEKBO survey intersects the Sgr plane at an angle and best overlaps at $\alpha \approx 18$h (i.e., behind the Galactic Center)
and $\alpha \approx 5$h.  In the regions traced by the Prior et al. (2009) subsample we confirm that the RRL tend to lie
$\sim 10^{\circ} - 15^{\circ}$ away from the bulk of the Sgr stream.  It is perhaps unsurprising then that the distances
(typically 16-21 kpc) and radial velocities of the RRL do not obviously match any particular branch of the Sgr stream.
While Prior et al. (2009) discussed a possible match between some RRL and trailing Sgr tidal debris in the oblate-halo model
of LJM05, no such match obviously exists for the updated model presented here.

We note that the distance ($\sim 20$ kpc) and radial velocity range ($v_{\rm GSR} \approx -200$ - $+200$ \kms)
of the $\Lambda_{\odot} \approx 10^{\circ}-40^{\circ}$
sample of Prior et al. (2009) is similar to that found for a sample of red clump stars by Majewski et al. (1999; crosses in Figure \ref{OtherObs.fig}).
These red clump stars lie near the  angular position of the L1 stream, and LJM05 discussed the possibility that their large spread in radial velocities
may indicate the range of velocities occupied by the Sgr stream at this longitude.  Given the similarities in distance and radial velocity
of the Prior
et al. (2009) and Majewski et al. (1999) samples, it is at least likely that they trace the same Galactic substructure.  
If this substructure is confirmed to be the Sgr stream, it will suggest that
the L1 stream at $\Lambda_{\odot} \sim 25^{\circ}$ has a greater angular width, smaller distance, and larger spread in radial velocities than
predicted by the current $N$-body model.

\subsection{Subgiant Populations in the Leading Tidal Stream}

Keller (2009) has recently described a population of subgiant stars
in the range $\alpha \approx 120^{\circ} - 180^{\circ}$ from the SDSS that may correspond to the L1 stream of Sgr.
While this paper gives insufficient information to include these stars in Figure \ref{OtherObs.fig}, we note that the $\Lambda_{\odot}$ vs distance
trend illustrated in Figure 7 of Keller (2009) is reproduced by the revised $N$-body model to within $\sim 1$ kpc, in contrast to previous
$N$-body models of the Sgr stream (e.g., LJM05).
Further characterization of this subgiant population may therefore provide even greater insight into the nature of the 
Sgr leading tidal stream.


\section{Summary}
\label{summary.sec}

We have presented the first numerical model of the Sgr --- Milky Way system that is capable of simultaneously satisfying the majority of major
constraints on the angular positions, distances, and radial velocities of the dynamically young tidal debris streams.
In particular, we have resolved the so-called ``halo conundrum'', whereby models were previously incapable of fitting both
the angular position and distance/radial velocity trends of leading tidal debris.
The publically-available $N$-body model also makes concrete predictions
for the location and kinematics of older tidal debris not yet conclusively observed in the Galactic halo.  Future observational evidence
for the presence/absence of such streams will place strong constraints on the total interaction time of the dSph with the Milky Way.

The key ingredient in the success of this model is our introduction of a non-axisymmetric component to the Galactic gravitational potential,
in the form of  a nearly oblate dark matter halo with axial ratios $1.00/1.36/1.38$ whose minor 
axis lies within the plane of the Galactic disk plane within $7^{\circ}$ of the line connecting the Sun and the Galactic center.
While  this inclined-halo explanation permits the model to reproduce the observational data, it is not strongly motivated
within the current CDM paradigm, and may simply serve as a numerical crutch
that mimics the effect of some as-yet unidentified alternative (such as Sgr orbital evolution or the gravitational influence
of the Magellanic Clouds).

Our $N$-body model suggests that Sgr has been orbiting in the gravitational potential of the Milky Way for at least 3 Gyr and consists
of a remaining bound mass $M_{\rm Sgr} = 2.5^{+1.3}_{-1.0}  \times 10^8 M_{\odot}$.  The total space velocity of Sgr is predicted to be $v_{\rm Sgr} = 304$ \kms,
which at a distance $D_{\rm Sgr} = 28$ kpc corresponds to an apparent proper motion 
of $(\mu_{\alpha} \textrm{cos} \delta, \mu_{\delta} ) = (-2.45, -1.30)$ mas yr$^{-1}$ and a radial velocity of 142 \kms in the heliocentric rest frame.
In contrast to some previous models, the Sgr stream is not expected to approach the solar position to closer than $\sim 13$ kpc.
We have additionally demonstrated that it is possible to understand the observed evolution of the MDF along the Sgr tidal streams
in terms of the dynamical superposition of stars formed in multiple bursts of star formation and torn from the satellite at different times,
suggesting that the chemical abundances of stars in Sgr at the present day may
be significantly different than the abundances of those contributed to the Galactic stellar halo.

Despite our success in fitting the strongest observational constraints on the primary wraps of the leading and trailing tidal arms, it is uncertain
what gives rise to the low surface brightness, high-declination `bifurcation' of the stream that parallels the leading arm at $\sim 15^{\circ}$ higher declination
in the SDSS footprint.  
We postulate that substructure or anisotropic internal dynamics within the progenitor satellite may cause the observed bifurcation, indicating
that future numerical models of the Sgr stream must 
explore the influence of the internal dynamics of the progenitor on the subsequently formed tidal tails.  Such models must include realistic treatments
of the dark and baryonic material as separate components and explore both rotation and orbital anisotropy of the visible matter.
Moreover, the effects of evolution in the Sgr orbit and the peturbative gravitational
influence of Milky Way substructures such as the LMC must be investigated in greater detail.

\acknowledgements
DRL gratefully acknowledges productive conversations with Kathryn Johnston, Lars Hernquist, and Nate McCrady.
The authors also thank the referee for constructive suggestions which improved this manuscript.
Support for this work was provided by NASA through Hubble Fellowship grant \# HF-01221.01
awarded by the Space Telescope Science Institute, which is operated by the Association of Universities for Research in Astronomy, Inc., for NASA, under contract NAS 5-26555.  SRM acknowledges support from NSF grant AST-0807945 as well as the
{\it SIM Lite} key project {\it Taking Measure of
the Milky Way} under NASA/JPL contract 1228235.

\end{document}